%% file: paper.tex
\documentclass{aa}  

\usepackage{graphicx}
\usepackage{txfonts} 
\usepackage{natbib} 
\usepackage{xcolor}
\usepackage{amsmath} 
\usepackage{epstopdf}
\usepackage{linenoaa} 
\usepackage[colorlinks=true, citecolor=blue]{hyperref}

\begin{document}

\title{Dynamical conditions for tilting Neptune during planetary migration}
\subtitle{}

\author{
  Andrea Moretti\inst{1,2}
  \and
  Giacomo Lari\inst{2}
}

\institute{
  LTE, CNRS, Observatoire de Paris,
PSL University, Sorbonne Université,
77 Avenue Denfert-Rochereau, 75014 Paris, France
  \and
  Department of Mathematics, University of Pisa, Largo Bruno Pontecorvo 5, 56127 Pisa, Italy\\
}

\date{Received xxx; accepted xxx}

\abstract
{Giant planets are expected to form with near-zero obliquities, so that their current tilts must be the result of some post-formation processes. Capture into spin-orbit resonances has been shown to provide a viable mechanism for explaining the current obliquities of Jupiter, Saturn, and Uranus.}
{We aim to determine whether specific configurations of a primordial satellite system of Neptune could have enabled the tilting of its spin axis during planetary migration.}
{We modeled Neptune's secular spin-axis dynamics, including the presence of a primordial satellite system that enhanced the planet's precession constant. We then performed numerical simulations to track Neptune's obliquity evolution during planetary migration, exploring the conditions on the primordial satellite system that could have made Neptune trigger a spin-orbit resonance.}
{We identified an open set in the parameter space of the moon system that led to a successful tilting of Neptune. In particular, the total mass of the primordial system must have been $\geq 4\times 10^{-4}\,M_N$. A representative configuration consists of an inner satellite of mass $5\times 10 ^{-5}\,M_N$ and semi-major axis $35\,R_N$, and an outer satellite of mass $3.5\times 10 ^{-4}\,M_N$ and semi-major axis $100\,R_N$. Such a system could have survived during the unstable phase of the planetary migration if the gravitational encounters between Neptune and Uranus were not too close. The eventual ejection of the outer moon, possibly due to the encounter with a planetesimal, froze Neptune's obliquity to its current value.}
{Neptune's current obliquity can be attained during planetary migration if the planet initially possessed a satellite system with a suitable architecture. The required satellite properties are not too far from theoretical predictions and are consistent with recent scenarios proposed for the tilting of Uranus. Once the outer moon was ejected, the system matched a configuration suitable for the following capture of Triton and its fast evolution toward its current orbit.}


\maketitle
\nolinenumbers

\input{sections/01_Introduction.tex} 
\input{sections/02_SolarSModel.tex} 
\input{sections/03_NeptuneS.tex} 
\input{sections/04_Obliquity.tex} 
\input{sections/05_SatelliteSurvival}
\input{sections/06_Discussion}
\input{sections/07_Conclusions}

\begin{acknowledgements}
We thank Melaine Saillenfest for fruitful discussions. We are also grateful to the anonymous reviewer for their constructive comments and suggestions, which helped to improve the manuscript.
\end{acknowledgements}

\bibliographystyle{aa}
\bibliography{refs}

\end{document}

%% file: sections/01_Introduction.tex
\section{Introduction}\label{sec_intro}

Giant planets are expected to form with rotation axis nearly perpendicular to their orbital planes, resulting in obliquities close to zero \citep{Ward-Hamilton_2004,Rogoszinski-Hamilton_2020}. However, the current tilts of giant planets in our solar system exhibit a wide range of values: Jupiter has an obliquity of approximately $3.1^\circ$, Saturn around $26.7^\circ$, Uranus around $97.8^\circ$, and Neptune around $28.3^\circ$. This suggests that some dynamical mechanisms must have influenced the planets' spin-axis orientations after their formation.

The tilting of planets can occur through various processes and requires studying the complex interplay between their rotational and orbital dynamics. Different hypotheses have been proposed to explain the observed obliquities of giant planets, including giant impacts \citep{LISSAUER-SAFRONOV_1991,SLATTERY-etal_1992} and secular spin-orbit resonances \citep{Ward-Hamilton_2004,Ward-Canup_2006}. These mechanisms can operate at different stages of planetary evolution, and may involve also planets' moons and their orbital migration \citep{Boue-Laskar_2006,Saillenfest-eatal_2020,Saillenfest-etal_2021a}. The processes that led to the current obliquities of the giant planets are markers of their formation and evolutionary history and thus provide insights into the dynamics of planetary systems in general.

Among the ice giants, Uranus's obliquity have drawn most of attention because of its extremely large value. Nevertheless, understanding the origin of the current tilt of Neptune is equally important to constrain the planet's evolution and make a comparison between the two planets. Current estimations place it approximately at $28.32^\circ$, while its spin-axis precession rate is $0.0134 \text{ "yr}^{-1}$ \citep{Jacobson_2009,Archinal-etal_2018}. 

A possible explanation for Neptune's tilt is a series of impacts with large planetesimals that probably occurred at the end of the accretion phase \citep{Parisi-DelValle_2011}. Such collisions could have delivered a substantial amount of angular momentum, ultimately tilting Neptune’s spin axis to its present state. Within this framework, differences in the impact histories of Uranus and Neptune could account for the qualitative differences in their satellite systems as well as their distinct internal heat fluxes \citep{Reinhardt-etal_2020}. Nevertheless, this scenario is not without difficulties, as discussed in previous works \citep{Rogoszinski-Hamilton_2020,Saillenfest-etal_2022}. More precisely, the notable similarities between Uranus and Neptune may instead point toward a more gradual and coherent evolutionary pathway, akin to that proposed for Jupiter and Saturn, rather than to stochastic late-stage collisional events.

A smooth tilting mechanism is provided by secular spin-orbit resonances. During its evolution, Neptune's spin axis could have been captured into resonance with one of the precessional frequencies of its orbit. Since evolution into resonance requires variations of some dynamical frequencies, capture into secular spin-orbit resonance can occur at an early stage of planets' history, e.g. during their orbital migration \citep{Ward-Hamilton_2004,Ward-Canup_2006}, or at a later stage, possibly because of the orbital expansion of their moons \citep{Saillenfest-eatal_2020,Saillenfest-etal_2021a,Wisdom-etal_2022}.

In this context, Saturn is a prominent example of how its current obliquity could be the result of a capture into spin-orbit resonance. Specifically, a resonance with the fundamental frequency $s_8$, which is generally associated with Neptune's nodal precession, is now the most accepted theory for explaining Saturn's tilt \citep{Ward-Hamilton_2004,Saillenfest-etal_2021a,Saillenfest-etal_2021b, Wisdom-etal_2022}. It has been also proposed that Uranus's extreme obliquity could have been caused by a spin-orbit resonance, but this requires that the planet hosted in the past a more massive satellite system than it has today \citep{Boue-Laskar_2010,Saillenfest-etal_2022} or a massive circumplanetary disk \citep{Rogoszinski-Hamilton_2020}. 

Motivated by these results, we can wonder whether, during its evolution, Neptune encountered a secular spin-orbit resonance able to increase its obliquity to its present value. Applying this mechanism to Neptune is challenging because its large heliocentric distance and relatively low-mass present-day satellite system place it far from any strong first-order secular spin-orbit resonance (see Table~\ref{tab:obliquities}).

Neptune may, however, have hosted a more massive primordial satellite system in the past \citep{Mosqueira-Estrada_2003}. The presence of ancient moons could have increased Neptune's precession constant, potentially allowing the planet to be captured into resonance during the orbital migration of the giant planets. The number, size, and orbital configuration of these satellites remain unknown, but some constraints can be derived from expected formation processes and comparisons with other satellite systems \citep{Mosqueira-Estrada_2003,Canup-Ward_2006,Rufu-Canup_2017}.

Furthermore, the transition from the primordial satellite system to the current one necessarily involves the capture of Triton, which is the current largest satellite of Neptune and presents a peculiar retrograde orbit. As of today, there is great uncertainty about the timing and exact capture scenario of Triton, with different proposed mechanisms \citep{Agnor-Hamilton_2006,Nesvorny-etal_2007,Nogueira-etal_2011, Rufu-Canup_2017}.

Different capture scenarios and orbital evolution of Triton can alter the history of Neptune's obliquity. In a recent study, \citet{Gomes_2026} showed that an initial highly eccentric orbit of Triton could have made Neptune increase its obliquity. More precisely, a large orbital eccentricity of the moon would have enhanced greatly the precession frequency of Neptune's spin-axis, and through the following orbital circularization Neptune would have crossed the spin-orbit resonance with $s_8$. In order to make the capture into resonance possible and then increase the obliquity of the planet up to its current value, \citet{Gomes_2026} assumed for Triton a timescale of circularization larger than $100$ Myrs, driven by tides alone, so that the tilting of Neptune occurred from a few tens of million years up to a few billion years after the capture of the moon. Such an evolution is in agreement with \citet{Nogueira-etal_2011}, which studied Triton's capture through binary exchange in the context of the Nice model, followed by long-term tidal evolution. However, in order to preserve the outer moon Nereid, it is necessary that Triton was captured very early, shortly after the final stages of Neptune's formation.

On the contrary, \citet{Rufu-Canup_2017} investigated a relatively late capture of Triton, placed during the final stages of planetary migration, when Neptune was traversing a planetesimal disk extending from approximately $20$ to $30$ au. As shown by \citet{CUK-GLADMAN_2005}, Triton's capture would have destabilized the primordial satellites of Neptune, and interaction with the resulting debris would have reduced Triton's semi-major axis and eccentricity to its current values in just $10^5$ years, a timescale much shorter than the one dictated by tidal dissipation. In this context, \citet{Rufu-Canup_2017} considered a primordial satellite system with a total mass between $0.3\times10^{-4}$ and $3.3\times10^{-4}$ Neptune masses. They found that the capture of Triton and following interaction with the moons can reproduce the current configuration of the system, with a large probability of success when considering the lower range of the total mass.

The rapid circularization of Triton's orbit permits the survival of Nereid, but it prevents Neptune's capture into spin-orbit resonance driven by the moon's orbital evolution. However, as already described, the presence of a primordial satellite could have allowed Neptune to trigger the spin-orbit resonance. The dynamical mechanism proposed by \citet{Gomes_2026} places the tilting of Neptune after the orbital migration of the planet, possibly even billions of years after the formation of Neptune. In this paper, instead, we investigate whether Neptune's present-day obliquity could have been generated through a secular spin-orbit resonance during the orbital migration of the giant planets. In particular, we explore the dynamical conditions to be imposed on the primordial satellite system that would have allowed Neptune to encounter such a resonance, increasing the planet's obliquity to its current value of $28.3^\circ$.

The structure of the paper is as follows: in Sect.~\ref{sec_2}, we describe the dynamical model and the methods used to simulate Neptune's obliquity evolution. In Sect.~\ref{sec_3}, we explore the properties of the primordial satellite system that permit Neptune to encounter the spin-orbit resonance with $s_8$ during its orbital migration. In Sect.~\ref{sec_4}, we present the results of our numerical simulations and show possible evolution of the obliquity of Neptune. In Sect.~\ref{sec_5}, we analyze the stability of the primordial satellite system in the different phases of the evolution. In Sect.~\ref{sec_6}, we highlight the dynamical conditions under which Neptune is captured into resonance and its obliquity is excited up to its current value. In the same section, we discuss the eventual exit from the spin-orbit resonance and the connection with the following evolution of Triton. Finally, in Sect.~\ref{sec_conclusions}, we summarize our results.

\begin{table}
    \caption{Current eigenfrequencies of the solar system and precession constants of giant planets.}
    \label{tab:obliquities}
    \centering
    \setlength{\tabcolsep}{4pt}
    \renewcommand{\arraystretch}{1.3}
    \begin{tabular}{r r ||@{\hspace{12pt}} r r r}
    \hline\hline
    \multicolumn{2}{c}{$s_i (\text{''} \, \text{yr}^{-1})$} & & $\alpha_p (\text{''} \, \text{yr}^{-1})$ & $\alpha_{p,sat} (\text{''} \, \text{yr}^{-1})$ \\
    \hline
    $s_5$ & $0.0000$   & Jupiter & $0.8750$  & $2.5969$ \\
    $s_6$ & $-25.9731$ & Saturn  & $0.1847$  & $0.7539$ \\
    $s_7$ & $-2.9082$  & Uranus  & $0.0083$  & $0.0419$ \\
    $s_8$ & $-0.6783$  & Neptune & $0.0020$ & $0.0152$ \\
    \hline
    \end{tabular}
    \tablefoot{
    A first-order resonance may occur when the ratio between the precession constant of a planet $\alpha_p$ and an eigenfrequency $|s_i|$ is close to unity. In the table are reported both planet-only precession constant $\alpha_p$ and the enhanced value $\alpha_{p,sat}$ due to the presence of massive satellites. These values are computed using Eq.~\eqref{equation_precession_constant_complete}.}
\end{table}

%% file: sections/02_SolarSModel.tex
\section{Dynamics of the spin axis}\label{sec_2}

\subsection{Equations of Motion} \label{sec_eq}
The motion of the spin axis of a planet is governed by the solar torque acting on the planet's oblate shape. Moreover, orbital precession due to the gravitational attraction of the other planets introduces perturbations in the planet's orbital dynamics. In the case of a fixed orbit, the obliquity $\epsilon$ remains constant and the spin axis precesses with a constant rate
\begin{equation}
    \phi_{\alpha} = \alpha\cos\epsilon
\end{equation}
where $\alpha$ is called precession constant and is defined as
\begin{equation} \label{precession_constant}
    \alpha  = \frac{3}{2}\frac{\mathcal{G}M_\odot}{\omega a_\odot^3}\frac{J_2 }{\lambda} (1-e_\odot^2)^{-3/2}\,.
\end{equation}
In Eq.~\eqref{precession_constant}, $\mathcal{G}$ is the gravitational constant, $M_\odot$ is the mass of the Sun, $\omega$ the spin rate of the planet, $a_\odot$ and $e_\odot$ its semi-major axis and eccentricity, $J_2$ its quadrupole moment, and $\lambda$ its normalized polar moment of inertia. We take $J_2 = 3.536\times 10^{-3}$ \citep{Jacobson_2009}. The parameters $\lambda$ and $\omega$ of Neptune are not well constrained; we assume $\lambda = 0.23$, and $\omega$ such that the rotation period $T_\text{rot}= 16\, \text{h}\ 11\, \text{min}$. These values are in the range $\lambda\in [0.22,0.24.5]$ and $T_\text{rot}\in [16\,\text{h},17.5\,\text{h}]$ predicted by \cite{Podolak-Helled_2012} and \cite{Helled-Forney_2020}.

The equations of motion of the spin axis in an inertial reference system can be written as 
\begin{equation} \label{spin_axis_equation}
    \frac{d\textbf{s}}{dt} = \alpha (\textbf{s} \cdot \textbf{n}) ( \textbf{s} \times \textbf{n})\,,
\end{equation}
where $\textbf{s}$ is the unitary spin axis vector and $\textbf{n}$ the normal to the orbital plane. The angle between these two vectors is the obliquity $\epsilon$ of the planet. If we consider the orbital perturbations due to the gravitational attraction of the other planets, the vector $\textbf{n}$ and the obliquity $\epsilon$ are not fixed anymore, but they evolve with time and the dynamics becomes more complex.
 
\begin{figure}[ht!]
    \centering
    \includegraphics[width=0.5\textwidth]{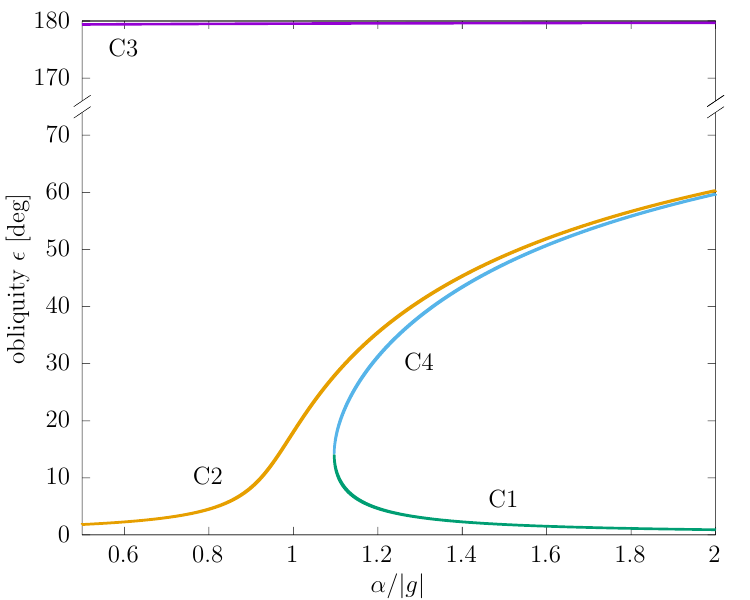}
    \caption{Obliquity at the Cassini states as a function of the ratio $\alpha/|g|$. In this plot, the inclination $I$ is set to $I_8(t_\text{end})$ (see Sect.~\ref{ssec_perturbation_planets}), which represents the inclination of Neptune's orbital plane when only the $s_8$ term is considered. The critical value of $\alpha/|g|$ for the appearance of Cassini States 1 and 4 is $\alpha/|g|_\text{crit} = (\sin^\frac{3}{2}I + \cos^\frac{3}{2}I)^ \frac{2}{3}\simeq 1.1$.}
    \label{fig_Cassini_states}
\end{figure}

\begin{figure*}[t]
    \includegraphics[width=0.48\textwidth]{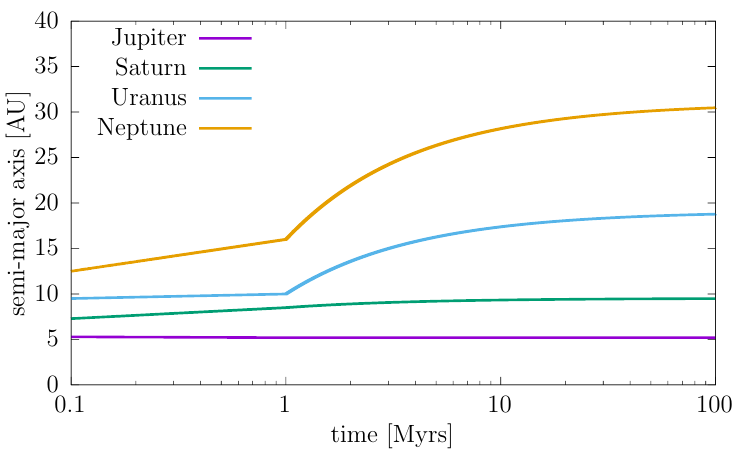}
    \qquad
    \includegraphics[width=0.48\textwidth]{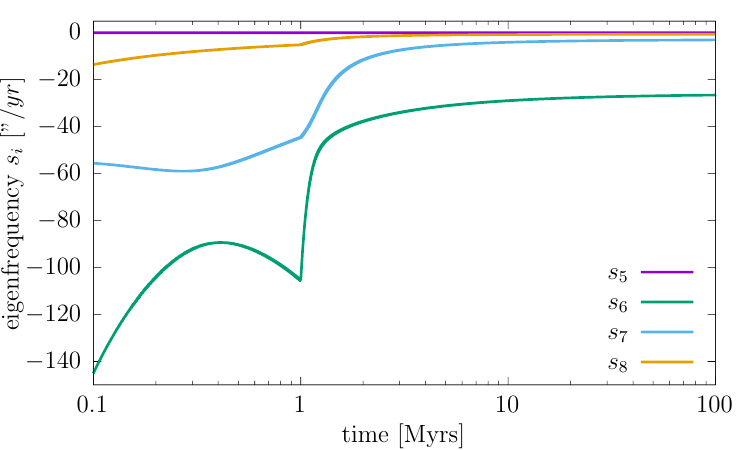}
    \includegraphics[width=0.48\textwidth]{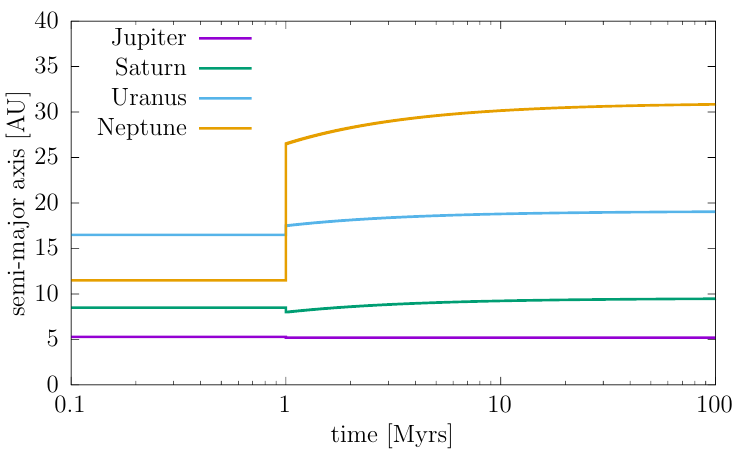}
    \qquad
    \includegraphics[width=0.48\textwidth]{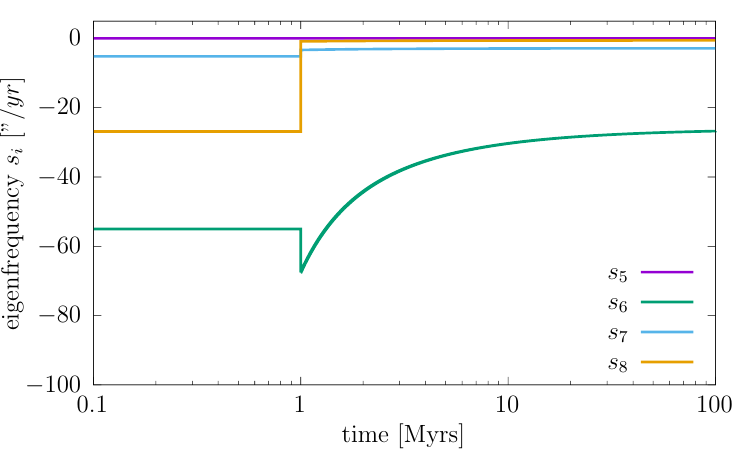}
    \caption{Examples of orbital evolution during the planetary migration. Top panel: variation of the giant planets' semi-major axes (left) and solar system eigenfrequencies (right) during a four-planet smooth Nice model simulation. Bottom panel: variation of the giant planets' semi-major axes (left) and solar system eigenfrequencies (right) during a four-planet Nice model simulation with a planet-planet scattering phase. The evolution of semi-major axes are taken from \cite{Nesvorny-Morbidelli_2012} and \cite{Tsiganis-etal_2005}, while the eigenfrequencies are computed as described in Sect.~\ref{ssec_perturbation_planets}.}
    \label{fig:semiassi_4}
\end{figure*}

\subsection{Cassini states}\label{ssec_cassini}
We first consider the case in which $\textbf{n}$ precesses uniformly around a fixed axis $\textbf{k}$ (the total angular momentum of the planetary system) with constant inclination $I$ and frequency $g$. This model, first studied by \cite{Colombo_1966} and later by \cite{Henrard-Murigande_1987}, provides a simplified framework that highlights the crucial role of spin-orbit resonances. In a reference frame that co-precesses with the orbital plane (taking $\hat{\textbf{e}}_3$ coincident with $\textbf{n}$), the equations of motion can be written as
\begin{equation}
    \frac{d\textbf{s}}{dt} = \alpha (\textbf{s} \cdot \hat{\textbf{e}}_3) (\textbf{s} \times \hat{\textbf{e}}_3) + g (\textbf{s} \times \textbf{k})\,.
\end{equation}
In such a coordinate system,
\begin{equation}
    \textbf{s} = (-\sin(\epsilon)\sin(\psi), \sin(\epsilon)\cos(\psi), \cos(\epsilon))\,,
\end{equation}
where $\psi = \phi_\alpha - \phi_g$ can be identified as the resonant angle. The topology of the phase space strongly depends on the parameters $\alpha$ and $g$, and in particular on their ratio $\alpha/|g|$. For $\alpha/|g| \lesssim 1$, the phase space has two equilibrium points, called Cassini states 2 and 3, and is foliated by invariant tori. For $\alpha/|g| \gtrsim 1$, the phase space topology becomes richer: two new equilibrium points (C1 and C4) appear, and a separatrix emerges. Libration around Cassini state 2 is then possible. In \figurename~\ref{fig_Cassini_states}, we show the evolution of obliquity at the Cassini states. We notice that the equilibrium point C2 drifts away from $\epsilon=0$ as $\alpha/|g|$ increases.

In a scenario in which $\alpha/|g|$ varies slowly over time, the spin axis can be captured into resonance around Cassini state 2. The probability of capture depends on the initial value of the obliquity and the rate of change of $\alpha/|g|$. Once captured into resonance, the obliquity can increase significantly as $\alpha/|g|$ continues to evolve. Differently from kinematic impacts, this kind of evolution leads to a smooth tilting of the planet. 

\subsection{Perturbation of the planets}\label{ssec_perturbation_planets}
In a more general scenario, the orbital plane of a planet does not precess uniformly. If quasi-periodic orbital motion is assumed, the evolution of the vector $\textbf{n}$ can be approximated as a superposition of periodic terms whose arguments are combinations of some fundamental frequencies \citep{Laskar_1990}. At first order in masses, and second order in both eccentricities and inclinations, these frequencies and the corresponding secular evolution of the planets can be obtained via the Laplace-Lagrange theory (see e.g. \citealp{Murray_Dermott_2000}). However, during planetary migration, the semi-major axes of the planets change and then the fundamental frequencies of the system are not constant. To take into account these variations, we consider the evolution of the semi-major axes of the planets as obtained by numerical simulations of the Nice model \citep{Tsiganis-etal_2005,Nesvorny-Morbidelli_2012}. From such evolution, we recover an expression of $\textbf{n}$ as a piece-wise secular Laplace-Lagrange system. We suppose that, at a given time step $t$ of the integration,
\begin{equation}
    \textbf{n} = (q(t), p(t), \cos I(t))\,,
\end{equation}
with
\begin{align}
    q(t) &= \sin I(t) \cos \Omega(t) = \sum_{j=5}^{8}I_{8,j}(t)\cos(s_j(t) t + \gamma_j(t))\,, \\ 
    p(t) &= \sin I(t) \sin \Omega(t) = \sum_{j=5}^{8}I_{8,j}(t)\sin(s_j(t) t + \gamma_j(t))\,,
\end{align}
where $I(t)$ and $\Omega(t)$ are Neptune's inclination and longitude of the node, respectively, while the sum is over all giant planets. In the above series, $s_j(t)$ are the eigenfrequencies of the solar system computed from planets' semi-major axes $a_j(t)$, and the parameters $I_{8,j}(t)$ and $\gamma_j(t)$ are the amplitudes and phases associated with the terms of the series. At each time step $t$, we evaluate the secular Laplace–Lagrange matrices using the current semi-major axes $a_j(t)$, and initialize $p(t)$ and $q(t)$ with the values obtained at the end of the preceding step. Evolutions are set to end at today's values of $p$ and $q$.

This model does not aim to describe accurately the whole system and it is not complete in any sense. For more realistic evolutions, full N-body numerical simulations should be used as in the Nice Model. Nevertheless, this approximation allows us to propagate only the spin-axis dynamics, while keeping the information from the orbital evolution of the planets. In the next section, we show some examples of the evolution of semi-major axes and eigenfrequencies for the giant planets of the solar system.

\subsection{Orbital solution}
The early history of the giant planets is still uncertain. In order to reconstruct it, constraints given by today's configuration of the solar system are used. However, they are not sufficient to unequivocally determine a single dynamical pathway. \cite{Tsiganis-etal_2005} first proposed a dynamical model of the early solar system, known as Nice model, that can explain many of its features, such as the current orbital configuration of the giant planets, the Late Heavy Bombardment, and the formation of the Kuiper Belt. The Nice model consists of a phase of smooth migration followed by an orbital instability, during which the planets could have experienced close encounters and significant changes in their orbits. So far, many variations of the Nice model have been proposed \citep{Morbidelli-etal_2007,Morbidelli-etal_2009,Nesvorny-Morbidelli_2012,Nesvonry_2011, Nesvorny_2018}, considering different initial conditions, number of planets, and timing of the instability. Each of these versions lead to a different migration history of the giant planets.

In this context, the wide range of possible migration scenarios would require testing our model against various semi-major axes evolution. However, this process would be computationally expensive. To address this, we limited our analysis to few representative cases. This decision was supported by several numerical integrations, which revealed similar system behavior when comparable histories were considered. Furthermore, as we will discuss in Sect.~\ref{sec_4}, the key processes of capture into resonance and obliquity increase occur in the last stages of planets' migration, therefore after the dynamical instability. While many orbital histories of the planets differ primarily in the timing and location of the instability, the final stages of migration exhibit similar characteristics. Two examples are given in \figurename~\ref{fig:semiassi_4}, which illustrates the semi-major axes and eigenfrequencies evolution for the giant planets of the solar system. The first scenario represents a smooth four-planet Nice model simulation \citep{Nesvorny-Morbidelli_2012}, while the other represents a four-planet Nice model simulation with a planet-planet scattering phase \citep{Tsiganis-etal_2005}. In both cases, the final stages of migration are similar, with Neptune approaching smoothly its current semi-major axis and the eigenfrequencies converging to constant values. Based on these observations, in our numerical integrations, we considered the evolution of the first scenario to analytically reconstruct the orbital evolution of Neptune (see Sect.~\ref{ssec_perturbation_planets}).

\subsection{Modeling of Neptune's primordial satellite system} \label{ssec_satellite_system}
In Sect.~\ref{sec_eq}, we presented the formula for the precession constant $\alpha$ not considering the contribution of the moons of the planet. The presence of a regular satellite system increases the value of $\alpha$. More precisely, far-away satellites - whose equilibrium orbital plane, called Laplace plane \citep{Tremaine-etal_2009}, is close to the planet's orbital plane - enhance the torque exerted by the Sun on the planet's equatorial bulge. Instead, close-in satellites - whose Laplace plane is close to the planet's equator - enhance the effective oblateness and rotational angular momentum of the planet.

Consequently, the effective precession constant in the presence of a regular satellite system becomes \citep{French-etal_1993}
\begin{equation}\label{equation_precession_constant_complete}
    \alpha  = \frac{3}{2}\frac{\mathcal{G}M_\odot}{\omega a_\odot^3 (1-e_\odot^2)^{3/2}}\frac{J_2'}{\lambda'}\,,
\end{equation}
where
\begin{align}
    J_2' &= J_2 + q= J_2  + \frac{1}{2}\sum_j \frac{m_j}{M_N} \frac{a_j^2}{R_N^2} \frac{\sin(2(\epsilon - i_j))}{\sin(2\epsilon)} \,,\label{equation_precession_constant_complete_J2} \\
    \lambda' &= \lambda + l = \lambda + \sum_j \frac{m_j}{M_N} \frac{a_j^2}{R_N^2} \frac{n_j}{\omega}\frac{\sin(\epsilon - i_j)}{\sin(\epsilon)}\,. \label{equation_precession_constant_complete_lambda}
\end{align}
In the formulas above, $m_j, a_j$, $n_j$, and $i_j$ are the mass, semi-major axis, mean motion, inclination of the Laplace plane with respect to the planet's equator of the $j$-th satellite, respectively. Parameter $R_N$ is Neptune's equatorial radius ($R_N=25\,225\,\text{km}$, \citealp{Jacobson_2009}), and $M_N$ is its mass ($M_N= 1.024\times 10^{26}\, \text{kg}$). The value of $i_j$ is computed applying the approximation of a massless satellite as in \cite{Tremaine-etal_2009} and \cite{Saillenfest-Lari_2021}.

These additional terms provide a significant boost to the precessional constant $\alpha$. Without moons, the value of $\alpha$ is generally not large enough to match any eigenfrequency $s_j$, as in the cases of Saturn and Uranus \citep{Ward-Hamilton_2004,Saillenfest-etal_2021a,Saillenfest-etal_2022}. More precisely, the indirect effect of the Sun's torque transmitted through the satellites is the dominant contribution to the spin-axis dynamics (see e.g. \citealp{Saillenfest-Lari_2021}). It is worth noting that Eqs.~\eqref{equation_precession_constant_complete_J2} and~\eqref{equation_precession_constant_complete_lambda} are valid for arbitrary values of the semi-major axis $a_j$, and they closely match the general precession solutions presented by \cite{Boue-Laskar_2006} in the case of a single satellite.

Nowadays, the only massive satellite of Neptune is Triton, which is not primordial, but it was captured at a later stage, probably during Neptune's migration \citep{Nesvorny-etal_2007,Rufu-Canup_2017}. Therefore, the primordial satellite system of Neptune could have been extremely different from how it looks now. \cite{Mosqueira-Estrada_2003} modeled the formation of the main satellites of Jupiter, Saturn, Uranus, and Neptune in a gaseous circumplanetary disk. According to their work, both Uranus and Neptune "lack" a massive satellite at around $0.1\cdot r_\text{H}$, where $r_\text{H}$ is the planets' Hill radius (equal to $57\,R_U$ for Uranus and $98\,R_N$ for Neptune). The presence of such a primordial satellite around Uranus is a plausible explanation for the large obliquity of the planet without invoking giant impacts \citep{Boue-Laskar_2010,Saillenfest-etal_2022}. Following the same line of reasoning, we consider that Neptune also possessed a primordial satellite system, which was later disrupted by some dynamical events. To explore this scenario, we conducted a comprehensive parameter study of the possible configurations of this system. To simplify the analysis, we considered the presence of only two moons: a close-in satellite modeled as an increase in the oblateness of Neptune, $J_2' = J_2 + \Delta J_2$, and a distant satellite with semi-major axis $a_s$ and mass $m_s$. The choice of the number of satellite is not restrictive. The effect of multiple close-in satellites can be approximated as a single satellite with an equivalent $\Delta J_2$, and, similarly, multiple far-away satellite can be approximated as a single satellite adjusting $m_s$. With these assumptions, Eq.~\eqref{equation_precession_constant_complete} becomes
\begin{equation}\label{eq_prec_satellite}
    \alpha  = \frac{3}{2}\frac{\mathcal{G}M_\odot}{\omega a_\odot^3 (1-e_\odot^2)^{3/2}}\frac{J_2 + \Delta J_2 + q}{\lambda + l}\,,
\end{equation}
where 
\begin{align}
    q &= \frac{1}{2} \frac{m_s}{M_p} \frac{a_s^2}{R_N^2} \frac{\sin(2(\epsilon - i_s))}{\sin(2\epsilon)}\,,\label{eq_q} \\
    l &= \frac{m_s}{M_p} \frac{a_s^2}{R_N^2} \frac{\sin(\epsilon - i_s)}{\sin(\epsilon)} \frac{n_s}{\omega}\,. \label{eq_l}
\end{align}
Here, $i_s$ is the inclination of the Laplace plane of the distant satellite with respect to the planet's equator, and $n_s$ its mean motion. In our simulations, we vary $\Delta J_2$, $a_s$, and $m_s$ to explore a wide range of possible ancient satellite configurations.

%% file: sections/03_NeptuneS.tex
\section{Capture into Resonance: the parameters space}\label{sec_3}

The problem we are studying can be viewed as a perturbed Colombo's Top \citep{Colombo_1966,Henrard-Murigande_1987}, where the resonant angle is $\psi = \phi_\alpha - \phi_{s_8}$. Without additional frequencies in Neptune's orbital expansion, this system would be integrable, as described in Sect.~\ref{ssec_cassini}. Despite this not being the case, we investigated whether the libration of $\psi$ dominates the spin-axis dynamics, treating it as a single-frequency Colombo's Top with perturbations deriving from the other orbital frequencies. We considered solely the resonance between $\phi_\alpha$ and $\phi_{s_8}$, as it is the most easily accessible. The ratios $\frac{\alpha}{|s_j|}$ for $j=6,7$, associated with the nodal precession of Saturn and Uranus, respectively, remain always far from unity, even including massive satellites in the Neptune system.

\subsection{Parameter's Manifold} \label{ssec_3_1}

From \figurename~\ref{fig_Cassini_states}, we deduce that a necessary condition for capture into resonance is $\frac{\alpha}{|s_8|}\geq 1$ at the end of planetary migration. We then define
\begin{equation}
    \label{eq:set}
    \mathcal{S} = \left\{ (\Delta J_2, a_s,m_s) \in \mathbb{R}^3 : \frac{\alpha(\Delta J_2,a_s, m_s; a_\odot^\text{end})}{|s_8|}\geq 1\right\}
\end{equation}
as the set of all possible primordial satellite systems which could allow for capture into resonance, as described in Sect.~\ref{ssec_satellite_system}. For computing Eq.~\eqref{eq_q}, we set $\epsilon=0$, while the values $a_\odot^\text{end}$ and $s_8$ are taken at the end of the orbital migration.

Given the nature of the parameters, some constraints should be applied. More precisely, for the close-in satellite, we considered $\Delta J_2 \in [0,0.05]$, where $\Delta J_2 = 0$ corresponds to the system in which just one far-away satellite is present, while $\Delta J_2 = 0.05$ is achieved by an inner satellite of mass $m = 5\times 10^{-5}\,M_N$ and semi-major axis $a = 50\,R_N$. For the far-away satellite, we let vary $a_s \in [50\,R_N,150\,R_N]$ and $m_s \in [10^{-4}\,M_N,\,10^{-3}M_N]$. The lower bound in the mass corresponds to values observed in Jupiter and Saturn's systems, while the upper bound is set to the absolute mass of a moon almost as large as Ganymede or Titan. Despite the distance $a_s$ appears to be quite large, the considered values are coherent with satellite formation models \citep{Mosqueira-Estrada_2003,Canup-Ward_2006}. In fact, although regular moons of Jupiter and Saturn are found way closer to the planet (Titan, for example, is at $20.4\,R_S$, while Callisto $26.5\,R_J$), we know that the right parameter to check is not $a_s/R_p$, but $a_s/r_\text{H}$, where $r_\text{H}$ is the Hill radius of the planet. Given Neptune's $r_\text{H} \simeq 4900\,R_N$, the far-away satellite we consider in our model is found in the range $[0.011,0.032]\, r_\text{H}$, which is comparable to the current positions of the satellites of Jupiter and Saturn. As an example, Titan is found at $0.02\,r_\text{H}^S$, where $r_\text{H}^S$ is Saturn's Hill radius, while Callisto at $0.035\,r_\text{H}^J$, where $r_\text{H}^J$ is Jupiter's Hill radius. For Neptune, these values would correspond to a distance of $93\,R_N$ and $162.5\,R_N$, respectively.
\begin{figure}[!ht]
    \centering
    \includegraphics[width=1.0\linewidth]{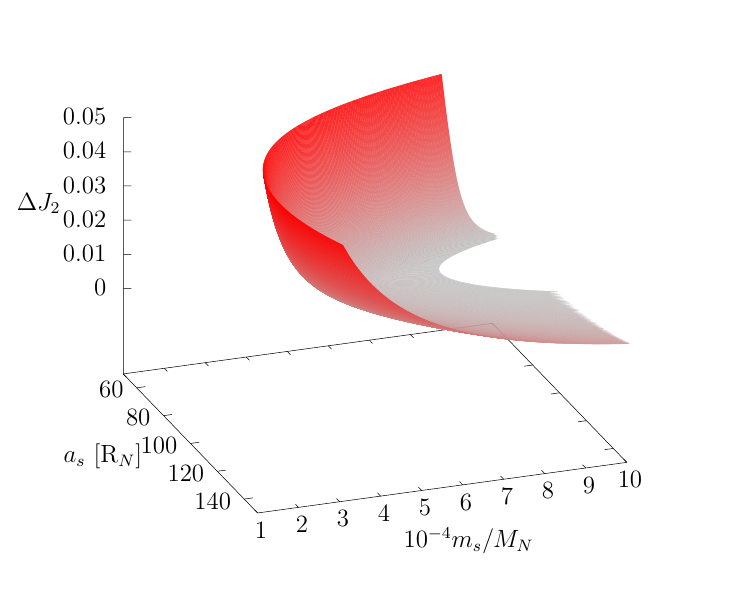}
    \caption{Boundary $\partial \mathcal{S}$ of the parameter's manifold. The region $\mathcal{S}$ is on its inside, toward greater values of the mass. }
     \label{fig_parameter_maifold}
\end{figure}

\subsection{Minimum mass}
\begin{figure}[!h]
    \centering
    \includegraphics[width=1.0\linewidth]{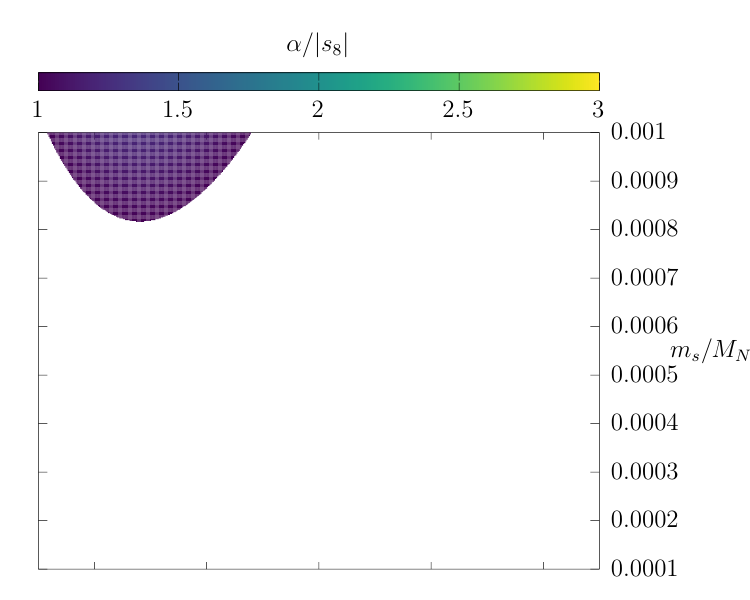}
    \qquad
    \includegraphics[width=1.0\linewidth]{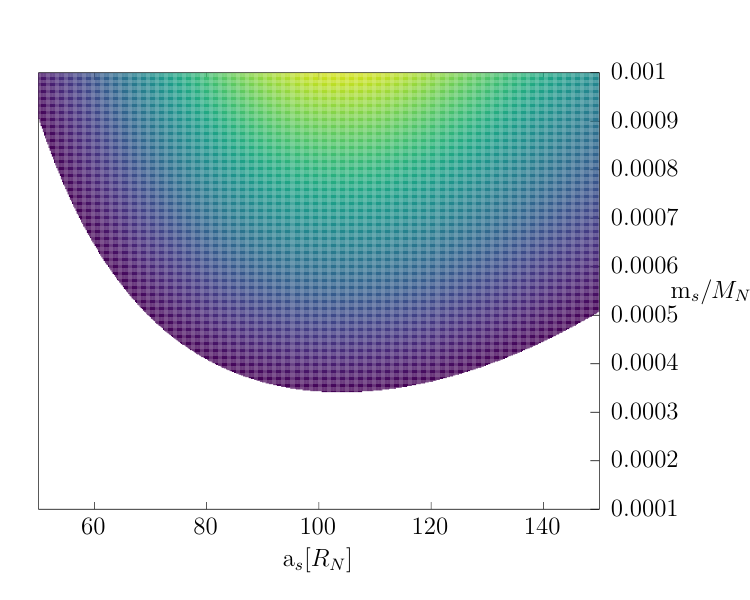}
    \caption{Sections of the parameter's manifold $\mathcal{S}$ at fixed values of $\Delta J_2$. Top panel: $\Delta J_2 = 0.0$. Bottom panel: $\Delta J_2 = 0.025$. The color bar indicates the value of $\alpha$ reached at the end of migration.}
    \label{parameters_curves}
\end{figure}

The parameter's set $\mathcal{S}$ turns out to be a $3-$dimensional closed analytic manifold (\figurename~\ref{fig_parameter_maifold}). Once fixed the values of $\Delta J_2$ and $a_s$, the condition defining $\mathcal{S}$ can be rewritten as
\begin{equation}
    m_s \geq \frac{J_2 + \Delta J_2 - B\lambda}{B\tilde{l}-\tilde{q}}\,,
\end{equation}
where
\begin{equation}
    B = \frac{|s_8|}{\frac{3}{2}\frac{\mathcal{G}M_\odot}{\omega a_\odot^3} (1-e_\odot^2)^{-3/2}}\,,
\end{equation}
and $\tilde{q} = q/m_s$, $\tilde{l} = l/m_s$. The latter quantities do not depend on $m_s$, since $q$ and $l$ are linear in $m_s$. This provides a straightforward method to analyze the parameter space and identify optimal parameters. From \figurename~\ref{parameters_curves}, which shows the $\Delta J_2 = 0$ section, we find that the minimum mass required for capture into resonance is $m_s \simeq 8\times 10^{-4}\,M_N$. While not impossible, this mass may seem somewhat too large as compared to the relative mass of regular moons formed around giant planets \citep{Canup-Ward_2006}. Moreover, smaller masses are preferable when studying the resonance break and Triton's capture \citep{Rufu-Canup_2017}. In contrast, examining $\Delta J_2 = 0.025$ section reveals a minimum mass of $m_s \simeq 3.8 \times 10^{-4}\,M_N$. Although still large compared to Neptune's current satellite system, this value is significantly more plausible \citep{Mosqueira-Estrada_2003,Rufu-Canup_2017}. The comparison between the two sections justifies therefore the choice of a double satellite system over a single one.

It is worth noting that, if Neptune was still undergoing thermal contraction at the time of planetary migration, its radius $R_N$ may have been larger than its present-day value. Since, the precession constant scales approximately as $\alpha \propto R_N$, its initial value could have been somehow larger. This would reduce the contribution needed by the satellite systems, thereby relaxing the mass and semi-major axis requirements for the inner and outer moons in our scenario. However, given the uncertainty surrounding Neptune's thermal evolution and the extent of any possible contraction, we adopt its present-day radius throughout our calculations.
 
\subsection{Strength of the satellite system}\label{ssec-strengh-sat}

Every triplet of parameters in $\mathcal{S}$ allows Neptune to encounter the resonance during the planetary migration, but not all of them are equally efficient in increasing its obliquity. From \figurename~\ref{fig_Cassini_states}, we observe that the Cassini state 2 moves to higher obliquities as $\alpha /|s_8|$ grows. Since we aim to reach current Neptune's obliquity of $\epsilon \simeq 28.3^\circ$, an appropriate range of $\alpha /|s_8|$ is needed. As we set $s_8$ at its current (final) value, we look for the effect of the parameters on the final value of $\alpha$. From Eqs.~\eqref{eq_prec_satellite}, \eqref{eq_q}, and \eqref{eq_l}, it follows that $\alpha$ grows as $\Delta J_2$ and $m_s$ increase. This is coherent with what we expect from the physics of the problem, since both parameters increase the total effective $J_2$ of the planet, which is directly proportional to $\alpha$. The dependence on $a_s$ is less trivial, as it affects $a_s^2$ terms in $q$ and $l$, but also the inclination of the Laplace plane $i_s$, and the orbital frequencies $n_s$. In the limit of small obliquity, \cite{Saillenfest-Lari_2021} obtained the expression
\begin{align}
    q_{\vert\epsilon = 0} &= \frac{1}{2} \frac{m_s}{M} \frac{r_M^2}{R_N^2} \left(\frac{a_s^2 / r_M^2}{1 + a_s^5 / r_M^5}\right)\,,  \label{q_eps_0}
    \\  l_{\vert\epsilon = 0} &= \frac{m_s}{M} \frac{r_M^2}{R_N^2} \left(\frac{a_s^2 / r_M^2}{1 + a_s^5 / r_M^5}\right)\frac{n_s}{\omega}\,, \label{l_eps_0}
\end{align}
where 
\begin{equation}
    r_M^5 = 2\frac{M_N}{M_\odot} (J_2+\Delta J_2) R_N^2a_\odot^3(1-e_\odot^2)^{3/2}\,.
\end{equation}
From Eq.~\eqref{q_eps_0}, they found that $q$ reaches a maximum for $a_{s,\text{max}}=(2/3)^{1/5}r_M$. Since the contribution of the $l$ term to the expression of $\alpha$ is weaker than that of $q$, the precession constant also reaches its maximum approximately at $a_s\simeq(2/3)^{1/5}r_M$. Therefore, for a given $m_s$, this value of $a_s$ approximately maximizes the ratio $\alpha/|s_8|$ (see \figurename~\ref{parameters_curves}).

The effect of $\Delta J_2$ is to shift $r_M$ toward larger values. Specifically,
\begin{equation}
\Delta\ln r_M=\ln\frac{r_M|_{J_2+\Delta J_2}}{r_M|_{J_2}}=\frac15\ln\left(\frac{J_2+\Delta J_2}{J_2}\right)\,.
\end{equation}

For $\Delta J_2=0.025$, this gives $\Delta\ln r_M\simeq0.42$, corresponding to $r_M|_{\Delta J_2 = 0.025}\simeq1.52\,r_M|_{\Delta J_2 = 0}$. Since the optimal satellite semi-major axis $a_{s,\text{max}}$ scales linearly with $r_M$, the maximum value of $q$, for a fixed satellite mass, satisfies
\begin{equation}
\Delta\ln q_{\max}=\ln\frac{q_{\max}|_{J_2 + \Delta J_2}}{q_{\max}|_{J_2}}=2\ln\frac{a_{s,\max}|_{J_2 + \Delta J_2}}{a_{s,\max}|_{J_2}}=2\Delta\ln r_M\,.
\end{equation}

Consequently, $q_{\max}|_{\Delta J_2 = 0.025}\simeq2.32\,q_{\max}|_{\Delta J_2 = 0}$. 
Therefore by shifting the optimal orbital distance outwards, the increase in $r_M$ is amplified through the quadratic dependence of $q_{\max}$ on $a_{s,\max}$. More precisely, the logarithmic increase in $q_{\max}$ is twice that of the optimal semi-major axis. As a first approximation, assuming that $q\gg J_2+\Delta J_2$ and $\lambda+l\simeq\lambda$, the same enhancement factor applies to $\alpha$. This explains why the $\Delta J_2=0.025$ region in \figurename~\ref{parameters_curves} is considerably larger than the $\Delta J_2=0$ region and how the presence of an inner satellite can substantially increase the maximum precession constant for Neptune.

\subsection{Numerical simulations setup}\label{ssec-stats}

Given different primordial satellite configurations, we expect different evolution of the spin axis of Neptune through the spin-orbit resonance with $s_8$. We therefore run numerical simulations considering different triplets of parameters $\Delta J_2$, $a_s$ and $m_s$ (see Sect.~\ref{ssec_3_1}) and we performed a statistical analysis of the outcomes. To this end, we sampled the parameter space $\mathcal{S}$ by selecting representative points that span the range of satellite masses, semi-major axes, and contributions to Neptune's $J_2$. More precisely, we took values of $a_s$ from $50\,R_N$ to $150\,R_N$ every $1\,R_N$, $\Delta J_2$ from $0.0$ to $0.05$ every $0.005$, and $m_s/M_N$ from $1.0\times 10^{-3}$ down to the boundary $\partial S$ in steps $1.0\times 10^{-5}$, slightly extending the sampling beyond $\partial S$ to account for the dependence of $\alpha$ on the obliquity during the evolution. 

Furthermore, for each selected configuration, we run $100$ simulations varying the initial obliquity (between $1$ and $5$ degrees) and resonant angle (between $0$ and $2\pi$), in order to assess the likelihood of capture into resonance.

For each simulation, we integrated Neptune's spin axis using the vector equation of motion given by Eq.~\eqref{spin_axis_equation}, with the precession constant $\alpha$ computed from Eq.~\eqref{eq_prec_satellite}, using the selected primordial satellite parameters. The evolution of Neptune's orbital plane is modeled as described in Sect.~\ref{ssec_perturbation_planets} taking into account the changes in semi-major axis during the planetary migration. We then tracked the evolution of Neptune's obliquity and resonant angle over a time span of $300$ Myr, consisting of $100$ Myr of migration followed by $200$ Myr of post-migration evolution. From the thousands of numerical simulations we run, we recorded the maximum obliquity of Neptune during the evolution, the probability of capture into resonance and the success in passing the current obliquity of $28.3^\circ$.

We note that in our setup, we did not integrate the orbital dynamics of Neptune's moons. We assumed that they have eccentricities close to zero, as expected for regular satellites, and fixed semi-major axes, as the considered timescale is too short to produce significant modification through tidal migration. Moreover, we assumed that the moons' inclinations follow the value given by their Laplace plane, which vary as the planet's obliquity increases. Since orbital oscillations about the Laplace equilibrium occur on a timescale shorter than that of the precession of Neptune’s spin axis, moons' average contribution can be modeled according to Eq.~\eqref{eq_prec_satellite} thorough the integration.

%% file: sections/04_Obliquity.tex
\section{Obliquity excitation}\label{sec_4}

As anticipated in Sect.~\ref{ssec-strengh-sat}, different triplets of parameters $(\Delta J_2, a_s, m_s)$ have different effects on Neptune's precession constant $\alpha$. This, in turn, influences the dynamics of Neptune's obliquity during the planetary migration. As we considered triplets that always allow to reach the resonance, the possible outcomes of the resonant encounter are that Neptune is captured into resonance and its obliquity gradually increase while the planet continues to migrate, or that it crosses the resonance without entering it and its obliquity undergoes an upward kick.

\begin{figure*}[t]
    \centering
    \includegraphics[width=0.49\textwidth]{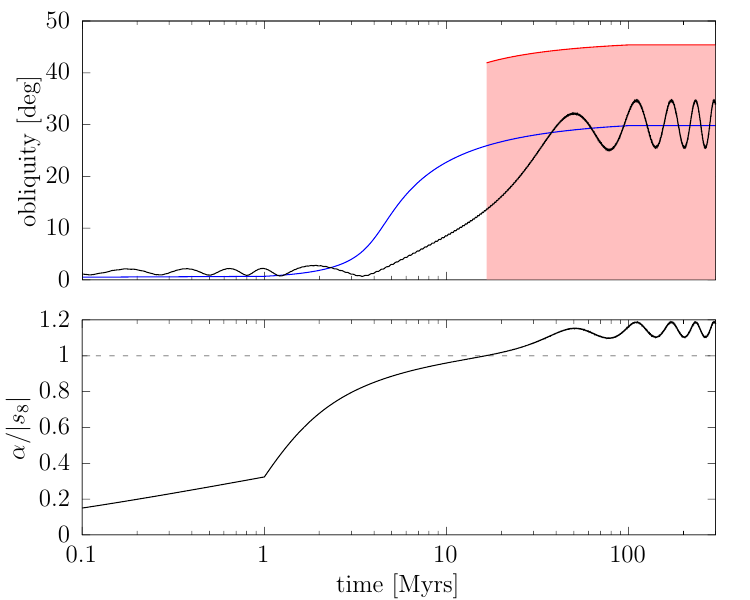}
    \includegraphics[width=0.49\textwidth]{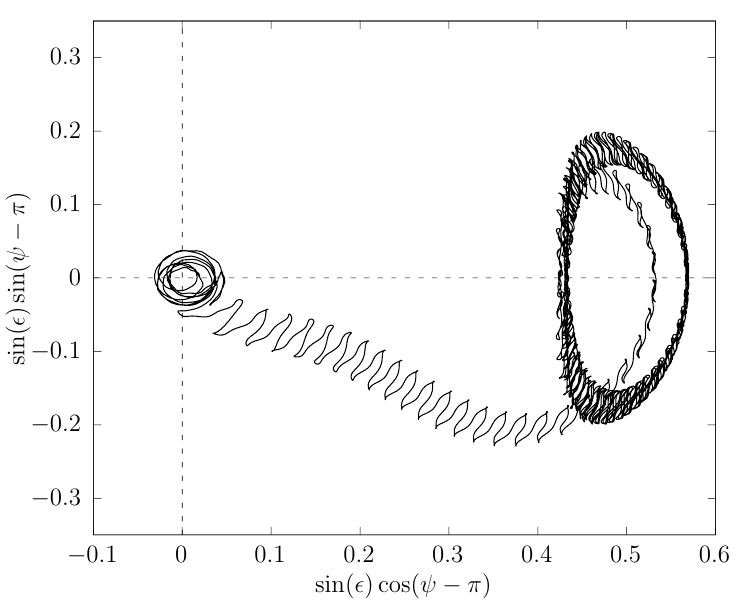}
    \caption{Evolution of the spin axis of Neptune during planetary migration for a representative case of successful excitation. Left top panel: evolution of the obliquity (black line); the blue line represents the obliquity at Cassini state 2, while the light red area is the region enclosed by the separatrix. For visual clarity, the Cassini state and the separatrix shown here are averaged values; their instantaneous values oscillate around the mean with an amplitude proportional to the oscillations of $\alpha/|s_8|$. Right panel: evolution of Neptune's spin-axis orientation on the unit sphere projected in the $(x,y)$ plane.}
    \label{fig:migration}
\end{figure*}

In the numerical simulations, we found both behaviors. In some cases, the obliquity of Neptune reached values of $28^\circ$ or higher, while in others did not. In this section, we present representative simulations of both successful and failed excitations, illustrating the dynamics involved in each scenario.
\begin{figure*}[t]
    \centering
    \includegraphics[width=0.48\textwidth]{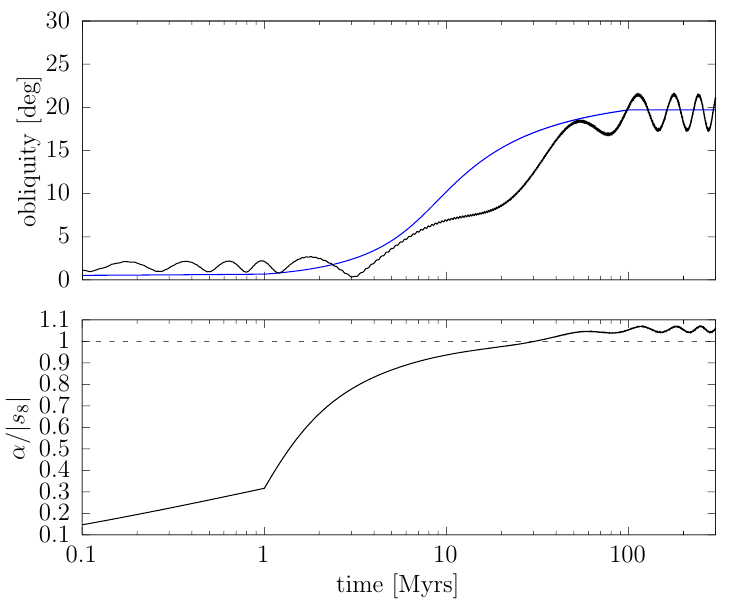}
    \includegraphics[width=0.48\textwidth]{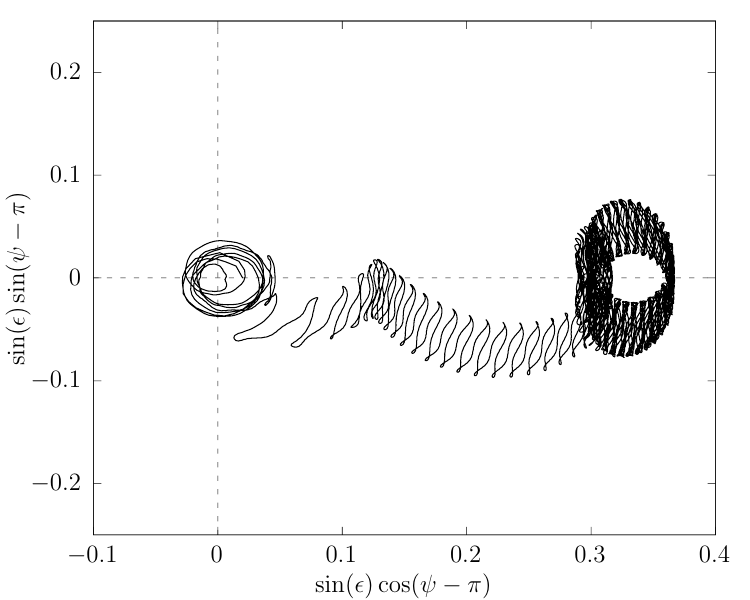}
    \qquad
    \includegraphics[width=0.48\textwidth]{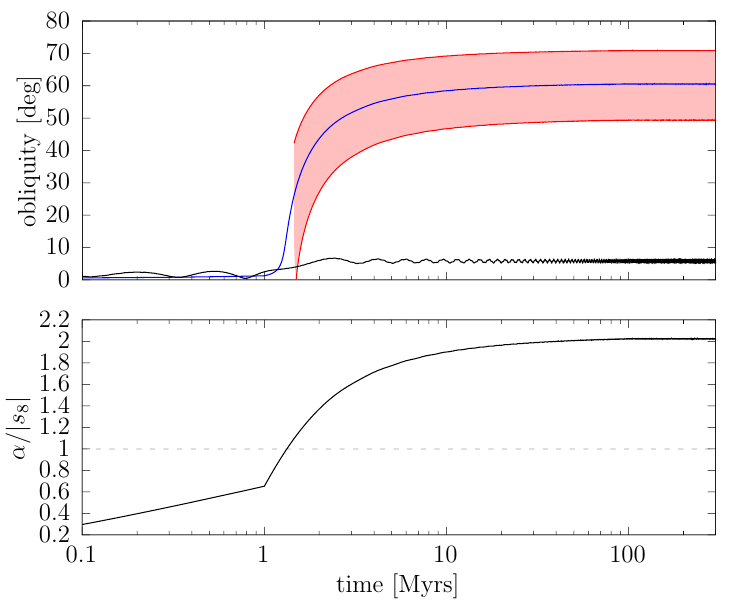}
    \includegraphics[width=0.48\textwidth]{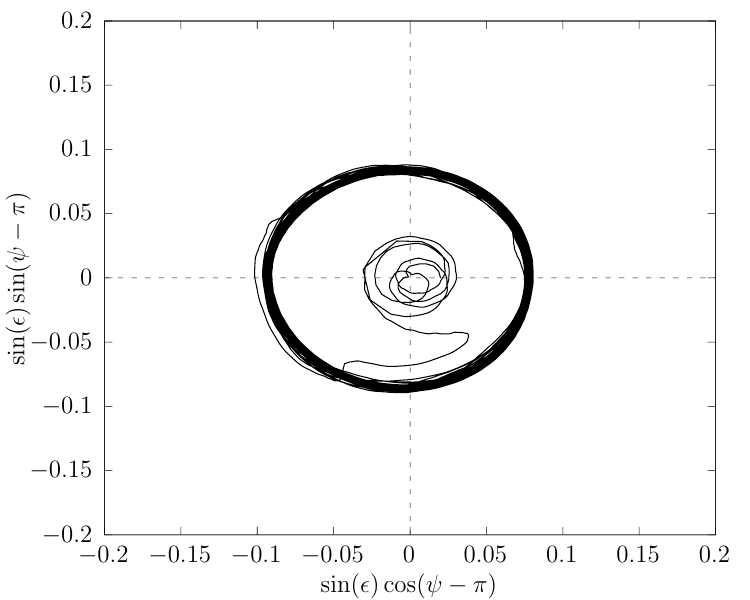}
    \caption{Top panel: evolution of Neptune's obliquity and spin-axis orientation during planetary migration for a representative case of shallow resonance. Bottom panel: evolution of Neptune's obliquity and spin-axis orientation during planetary migration for a representative case of failed capture into resonance. For visual clarity, the Cassini state and the separatrix shown here are averaged values; their instantaneous values oscillate around the mean with an amplitude proportional to the oscillations of $\alpha/|s_8|$. }
    \label{fig:migration_case1}
\end{figure*}
\subsection{When things go right: deep resonance}\label{ssec-right}
We start by presenting a case of a successful excitation of Neptune's obliquity. The primordial satellite system was set as $\Delta J_2 = 0.03$, $a_s = 100\,R_N$, and $m_s = 3.5 \times 10^{-4}\, M_N$. For this triplet of parameters, both probability of capture into resonance and success rate are $100\%$. In \figurename~\ref{fig:migration}, we reported the evolution of Neptune's obliquity and spin-axis orientation during migration for an initial obliquity around $1^\circ$. Initially, as both the ratio $\alpha / |s_8|$ and obliquity of Neptune are small, the spin-axis circulates around the north pole of the unity sphere. As migration proceeds, the ratio $\alpha / |s_8|$ increases, causing C2 to evolve toward higher obliquities. In this phase, the two timescales which dominate the dynamics are comparable. Indeed, the spin axis is neither librating or circulating around C2; nevertheless, the obliquity slowly increases, following the equilibrium point as it shifts. After approximately $20$ Myrs, $\alpha/|s_8|$ crosses the critical value, separatrix appears and the spin-axis is captured into resonance. From this point on, obliquity stabilizes and the spin-axis librates around C2. The value of $28.3^\circ$ is reached for the first time approximately after $35$ Myrs. After that, as obliquity oscillates, the same value is encountered periodically with a frequency given by the libration around the equilibrium point.

This evolution is typical of the successful runs we obtained. Different simulations are characterized by slightly different values of the final precession constant $\alpha$, which translates into different final obliquities, but the overall evolution is similar. In all cases, the obliquity of Neptune is excited up to its current value and the spin axis is captured into resonance, librating around C2. The main differences between simulations of this kind are the time at which the critical value of $\alpha / |s_8|$ is reached (which can happen anywhere between $20$ Myrs and $40$ Myrs), the value of maximum obliquity (which varies between $28^\circ$ and $40^\circ$), and the libration amplitude around C2. The latter is also influenced by the initial conditions of Neptune's spin-axis as it depends on the phase at the entrance of the resonance. In some cases, wide libration around C2 is observed, with the spin axis spanning a wide range of obliquity values during the libration. This is a consequence of the fast variation rate of $\alpha / |s_8|$ during migration, which makes the evolution through the resonance non-adiabatic. 

We note that at the end of Neptune's migration, the ratio $\alpha / |s_8|$ stops to increase and starts to oscillate (see \figurename~\ref{fig:migration}). This is due to the oscillation in obliquity, which affects the value of Neptune's precession constant through the term depending on the inclination of the outer moon (see Eqs.~\ref{eq_q} and~\ref{eq_l}).

\begin{figure*}[t]
    \centering
    \includegraphics[width=1.0\textwidth]
    {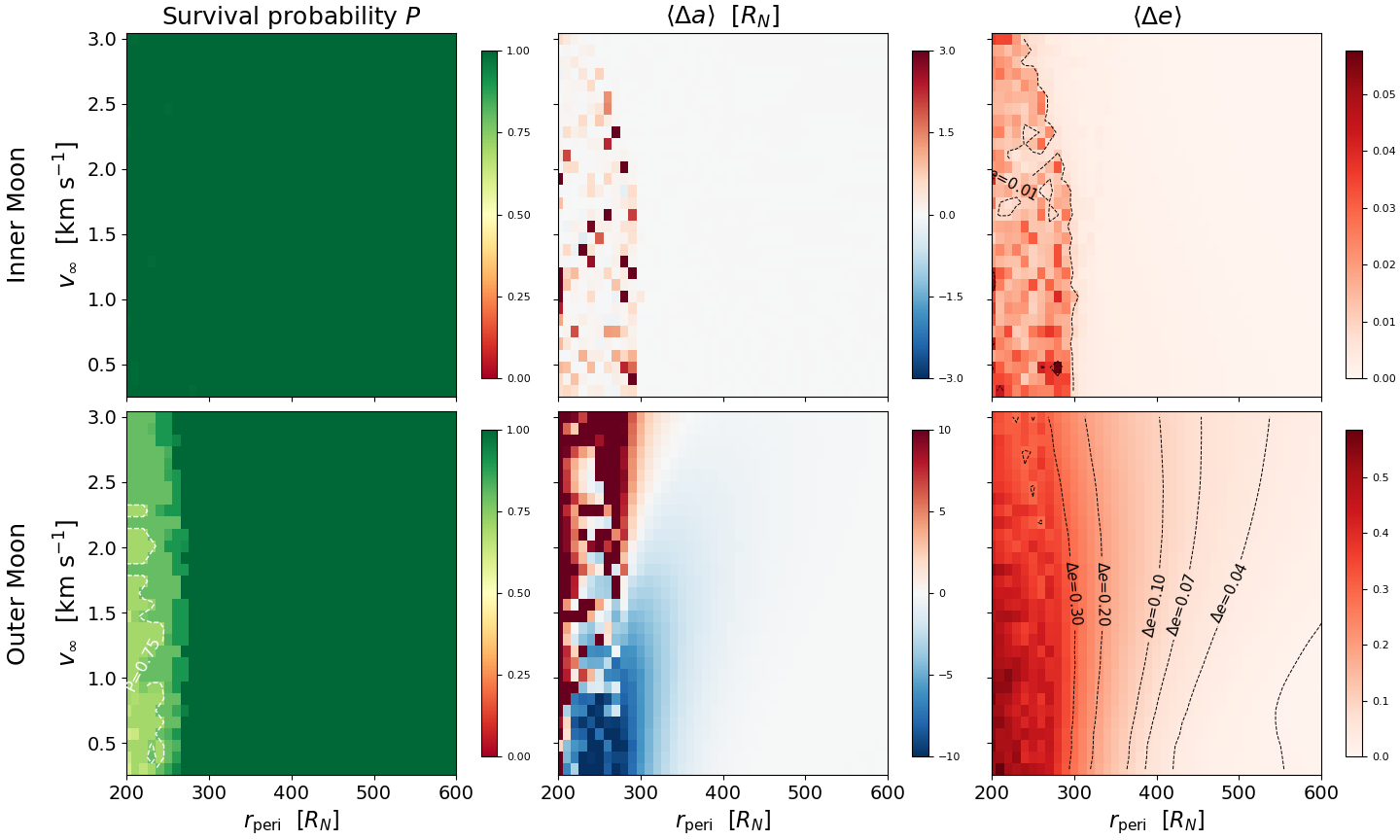}
    \caption{Survival probability and mean change in semi-major axis and eccentricity of the satellites as a function of the encounter parameters $r_{\text{peri}}$ and $v_\infty$. The left panels show the survival probability, while the middle and right panels show the mean change in semi-major axis and eccentricity, respectively. On the top, we report the results for the inner satellite; on the bottom, the results for the outer satellite.}
    \label{fig:encounter_survival}
\end{figure*}

\subsection{When things go wrong: failing capture or shallow resonance}
From our simulations, we identified two main factors that can prevent Neptune from reaching its current obliquity during migration: failure to capture into resonance and insufficient obliquity increase after capture. In \figurename~\ref{fig:migration_case1}, we reported two representative cases where these factors come into play.

In the first case (presented in the bottom panel of the figure) - parameters set to $a_s = 100\, R_N$, $m_s = 7.5 \times 10^{-4}\, M_N$, $\Delta J_2 = 0.03$ - Neptune is not captured into resonance. Here, the critical value of $\alpha/|s_8|$ is reached after approximately $1.5$ Myrs, a phase in which the migration timescale is too short compared to the libration period of the resonance, making capture impossible. As a result, the spin axis increases its obliquity because of the separatrix crossing, but ultimately it fails to be captured and settles on a circulating orbit throughout the simulation.

Different simulations show weaker or stronger kicks in obliquity, but for almost all of them Neptune fails to reach the desired value of $28.3^\circ$. In fact, only in very few simulations the obliquity of Neptune experiences a kick equal or larger than $28^\circ$, therefore we consider them statistically irrelevant.

In the second case (top panel) - parameters set to $a_s = 100\, R_N$, $m_s = 3.38 \times 10^{-4}\,M_N$, $\Delta J_2 = 0.03$ - Neptune is captured into resonance and its obliquity increases, but it never reaches $28.3^\circ$. Similarly to the successful scenario, the spin-axis starts the simulation circulating with small variations in obliquity. As migration proceeds, the ratio $\alpha / |s_8|$ increases, causing C2 to evolve toward higher obliquities. Here, the critical value of $\alpha/|s_8|$ for the appearance of the separatrix is never reached and Neptune's obliquity oscillates between $17^\circ$ and $22^\circ$.

This evolution is illustrative of what happens when the final value of $\alpha$ is not large enough. In simulations of this kind, the trajectory is found librating around Cassini State 2, with moderate amplitudes of oscillation around the equilibrium point. However, the final value of $\alpha / |s_8|$ does not permit Neptune to reach the requested obliquity of $28.3^\circ$.

%% file: sections/05_SatelliteSurvival.tex
\section{Stability of the satellite system}\label{sec_5}

The obliquity excitation mechanism observed in Sect.~\ref{sec_4} requires that the satellite system remained stable for, at least, $30$ Myr. This implies surviving both the giant planet instability phase, which typically occurred after $\sim 1$ Myr, and the subsequent phase of planetary separation, when Neptune started to cross the remnant planetesimal disk. Recently, \citet{Clement-etal_2026} investigated this problem in the context of violent Nice-model instabilities involving multiple close encounters among the giant planets. They followed the evolution of the Galilean satellites together with a primordial Uranian satellite system composed of its current moons plus six additional now-lost satellites. They found that, under such extreme conditions, the survival probability of either satellite system is below $15\%$, and only one realization out of 122 preserved both systems simultaneously. Although these results may suggest that such violent instabilities were unlikely, it is nevertheless important to assess the robustness of our proposed satellite configuration against similar perturbations.

\subsection{Encounters with giant planets}\label{sec_5b}
We started by investigating the effect of close encounters between the giant planets on the satellite system. In particular, we focused on the encounters between Neptune and Uranus, or an ice giant with an equivalent mass, since these are the ones that occur in most of the simulations \citep{Nesvorny_2018}. 

Using the numerical code \texttt{Rebound} \citep{REIN-LIU_2012}, we performed a series of numerical simulations of close encounters, considering a successful primordial system configuration from Sect.~\ref{sec_4}. In particular we investigated three different configuration of the outer moon: a satellite with mass of $1\times 10^{-3}\, M_N$ and semi-major axis of $ 55 \, R_N$, a satellite with mass of $3.5\times 10^{-4}\, M_N$ and $a_s = 100 \, R_N$ and finally one with mass of $4.5\times 10^{-4}\, M_N$ and $a_s = 145 \, R_N$. In all three configurations, the inner satellite was assigned a fixed mass of $5\times 10^{-5}\, M_N$ and semi-major axis of $a_s = 35\, R_N$.
The encounter was simulated as an hyperbolic passage of the ice giant at a minimum distance of $r_{\text{peri}}$ from Neptune, with a velocity at infinity of $v_\infty$ km/s. For simplicity, we adopted a small mutual inclination between the planetary and satellite orbital planes; this assumption is not restrictive, as the dominant perturbations depend primarily on the encounter distance and relative velocity. 

We tracked the evolution of the satellite system during the encounter and we assessed the stability of the system after such a dynamical event. In particular we let $r_{\text{peri}}$ and $v_\infty$ vary in the range $[100,600]\,R_N$  and $[0.5,3]$ km/s, respectively. For each pair of parameters, we sampled multiple initial satellite phases and computed the survival probability, together with the mean changes in semi-major axis and eccentricity. 

We found that the satellite system remained fully intact, with both moons surviving in all realizations, provided that $r_\text{peri}>160 \, R_N$ for the first configuration, $r_\text{peri}>270 \, R_N$ for the second configuration, and $r_\text{peri}>400 \, R_N$ for the last configuration. Such values show how too distant satellites are less likely to survive during the planetary instability, favoring a configuration with the outer moon at $a_s=100\, R_N$ over configurations with the outer moon at farther distances. Nevertheless, even non-disruptive encounters could have significantly altered the orbital architecture, particularly the semi-major axis and eccentricity of the outer satellite. Such changes may have accumulated over multiple encounters or modified Neptune's precession constant sufficiently to affect subsequent spin-orbit resonance capture.

\subsection{Encounter with planetesimals}\label{ssec_planetesimal}

The occurrence and strength of the encounters with giant planets during planetary migration depend on the setup and initial conditions of the solar system. There are models of the giant planet instability that predict a violent phase with many close encounters between the giant planets \citep{Clement-etal_2026}, while other models predict a more mild evolution with few or no close encounters \citep{Tsiganis-etal_2005, Nesvorny-etal_2007,Nesvorny-Morbidelli_2012}. Despite that, all models predict a phase of planetary separation, where Neptune and Uranus crossed the planetesimal disk, migrating outwards. This stage introduces a different threat to the satellite system: a large number of encounters with much smaller bodies.

To investigate this process, we performed a suite of simulations analogous to those described above, replacing the ice giant with a planetesimal of mass $m_p$ and considering an outer moon with $a_s = 100\, R_N$ and $m_s = 3.5 \times 10^{-4}\, M_N$. Because the planetesimal is substantially less massive than the ice giant, significant perturbations of the satellite system require much closer encounters. We therefore modeled each encounter as a hyperbolic flyby relative to Neptune, characterized by a Neptune-centered pericentric distance $r_{\rm peri}$ and asymptotic velocity $v_\infty$. For these simulations, we varied $m_p \in [10^{-4},10^{-3}]\,M_N$, $r_{\rm peri} \in [40, 100]\,R_N$ and $v_\infty \in [0.3, 3.5]\,{\rm km\,s^{-1}}$.

Unlike the previous set of simulations, the intruder's trajectory intersects the orbit of the outer moon. Consequently, the minimum separation between the two bodies, and hence the strongest instantaneous gravitational interaction, does not necessarily occur when the intruder is at pericenter relative to Neptune. We therefore recorded the true minimum distance $r_\text{min}$ between the planetesimal and the outer moon during each encounter. We then investigated how the ejection of the outer moon depends on $r_\text{min}$ and on the mass of the intruder $m_p$.

Our simulations showed that a single planetesimal encounter is much less disruptive than an encounter with another giant planet. The outer moon is ejected only for the most massive intruders, with $m_p \gtrsim 4\times 10^{-4}\,M_N$, and for very close planetesimal-outer moon encounters, with $r_\text{min} \lesssim 1\, R_N$. The ejection threshold depends strongly on the intruder’s mass and $v_\infty$. More massive and slower planetesimals can eject the outer satellite at larger encounter distances, whereas less massive or faster bodies require closer passages. For example, a planetesimal with $m_p = 10^{-3}\,M_N$ can eject the outer moon if it passes within $r_\text{min} \lesssim 1\,R_N$, while a planetesimal with $m_p = 4 \times 10^{-4}\,M_N$ requires a much closer approach of $r_\text{min} \lesssim 0.3\,R_N$ to achieve the same outcome. 

In addition to ejections caused by a single sufficiently close encounter, repeated planetesimal flybys may provide a further pathway for destabilizing the outer satellite. During planetary migration, the satellite system may undergo many of such encounters over a relatively short time interval, allowing their effects to accumulate. Indeed, substantial perturbations to the orbit of the outer satellite are observed even for encounters that do not result in immediate ejection. For example, a planetesimal with $m_p = 10^{-3}\,M_N$ passing at $r_{\min}\approx 3\,R_N$ can excite the satellite’s eccentricity to $e\approx 0.2$. A comparable eccentricity excitation can be produced by a lower-mass planetesimal with $m_p = 4\times10^{-4}\,M_N$, provided that it passes more closely, at $r_{\min}\approx 1\,R_N$. Such orbital perturbations may accumulate over successive encounters, potentially leading to the eventual ejection or long-term destabilization of the outer satellite.

In contrast, the inner satellite remains stable throughout the simulations. Its tighter gravitational binding to Neptune enables it to withstand perturbations from planetesimal encounters, even in cases where the outer satellite is ejected. This behavior is consistent with the outcomes of encounters involving the ice giant (see \figurename~\ref{fig:encounter_survival}) and indicates that a primordial inner satellite system could have survived the disruption of the outer moon.

%% file: sections/06_Discussion.tex
\section{Discussion}\label{sec_6}
\begin{figure*}[t]
    \centering
    \includegraphics[width=0.37\textwidth]{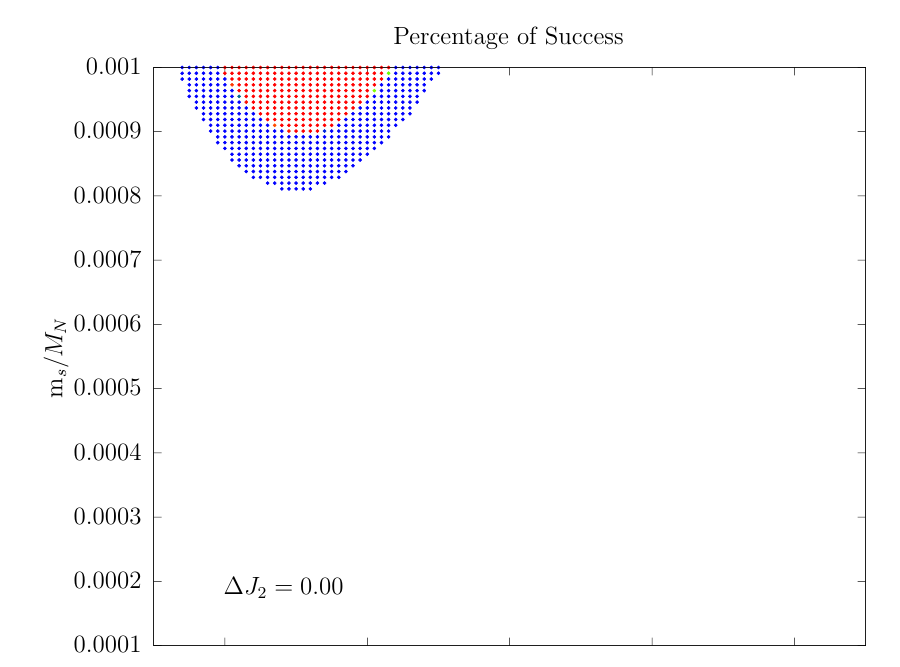}
    \includegraphics[width=0.31\textwidth]{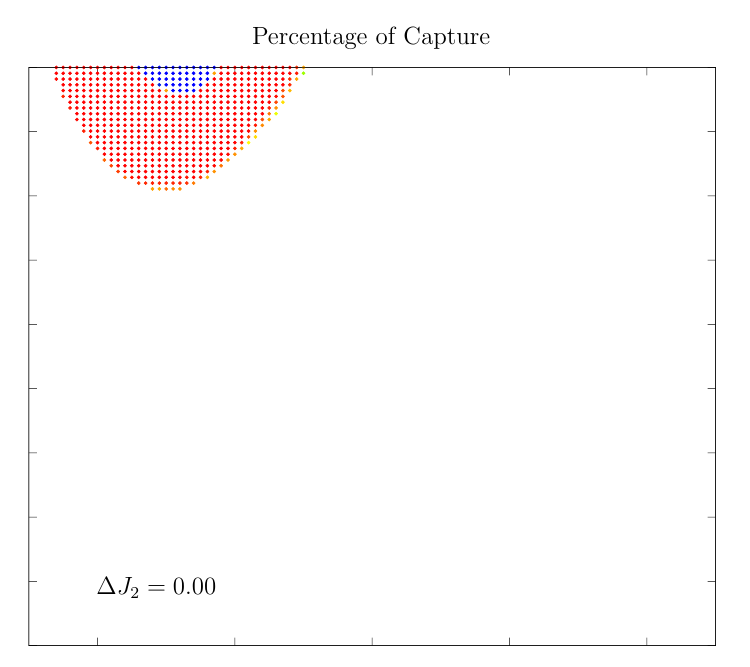}
    \includegraphics[width=0.31\textwidth]{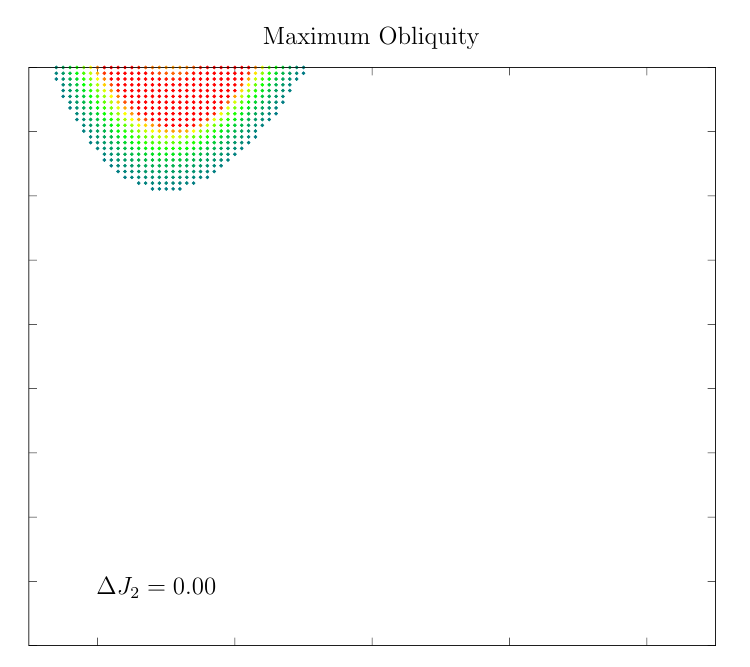}
    \qquad
    \includegraphics[width=0.37\textwidth]{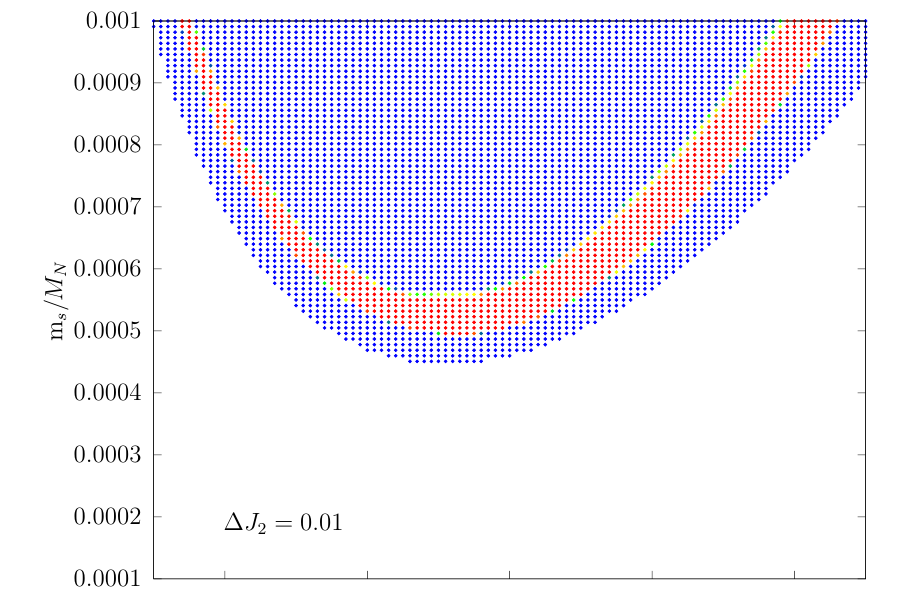}
    \includegraphics[width=0.31\textwidth]{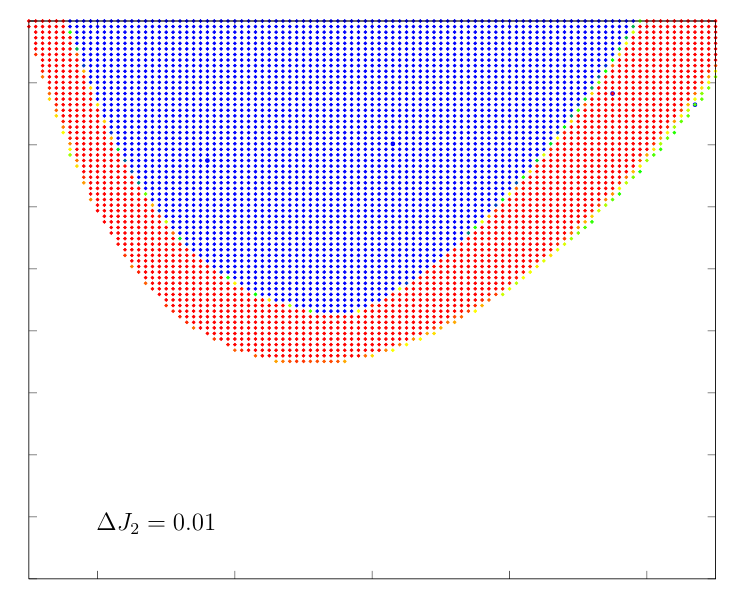}
    \includegraphics[width=0.31\textwidth]{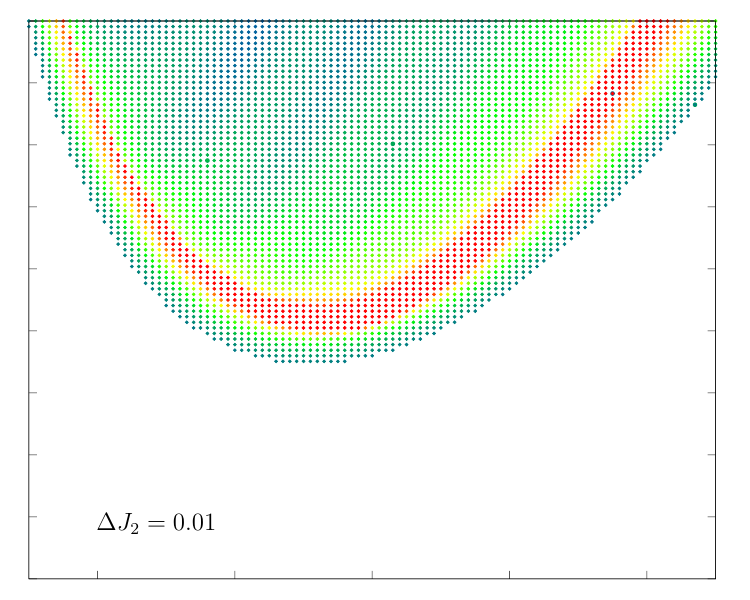}
    \qquad
    \includegraphics[width=0.37\textwidth]{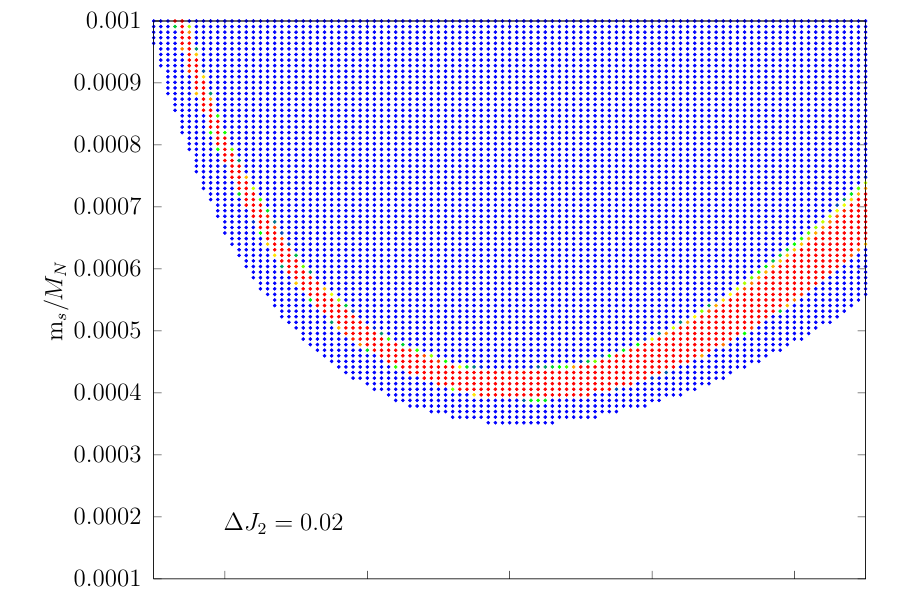}
    \includegraphics[width=0.31\textwidth]{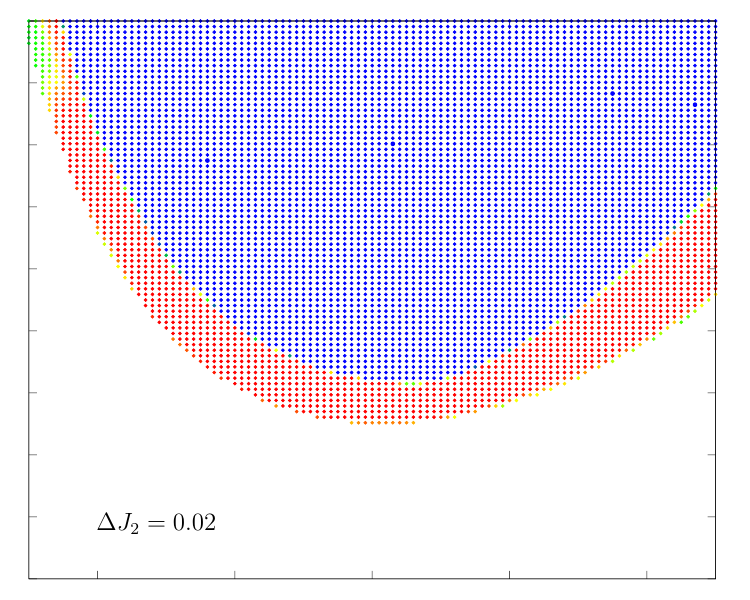}
    \includegraphics[width=0.31\textwidth]{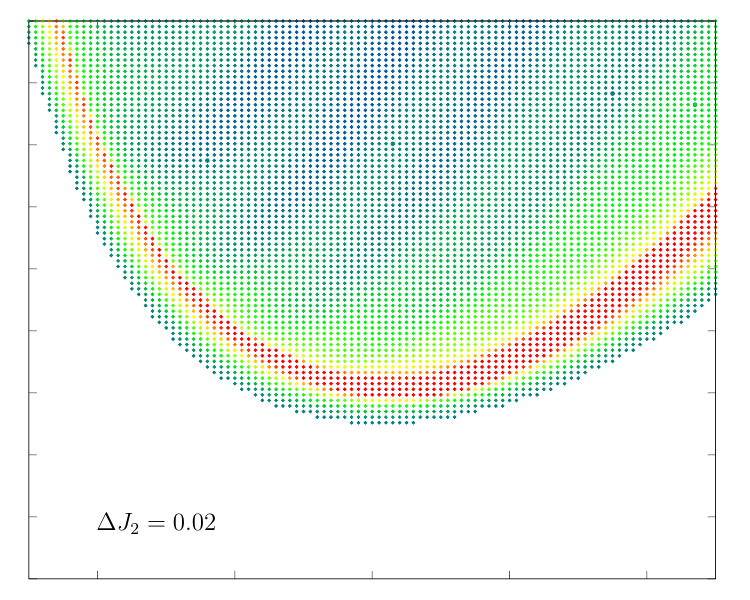}
    \qquad
    \includegraphics[width=0.37\textwidth]{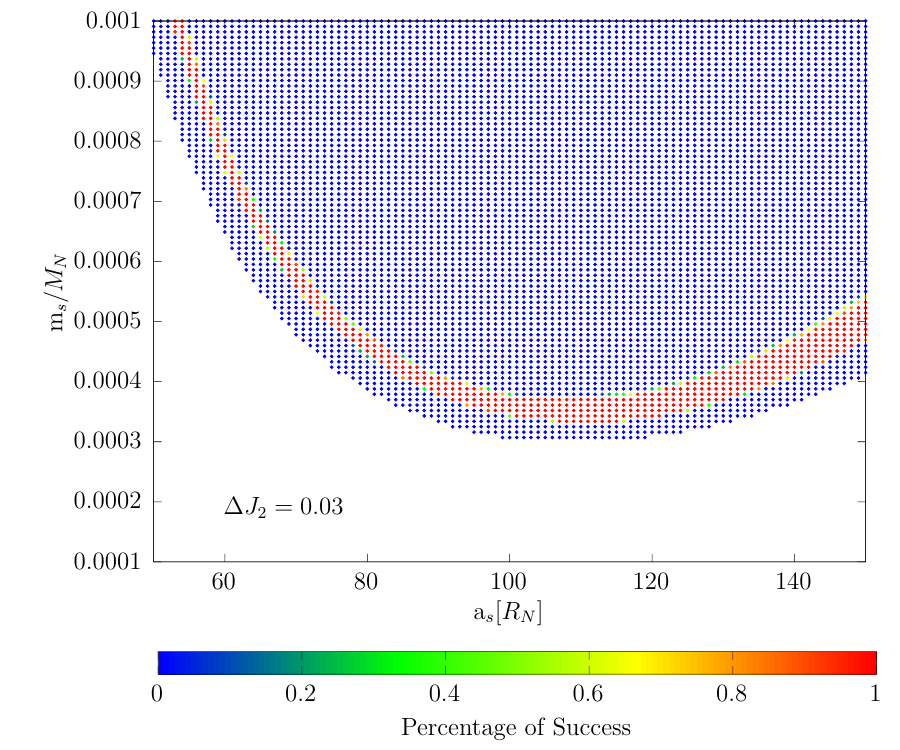}
    \includegraphics[width=0.31\textwidth]{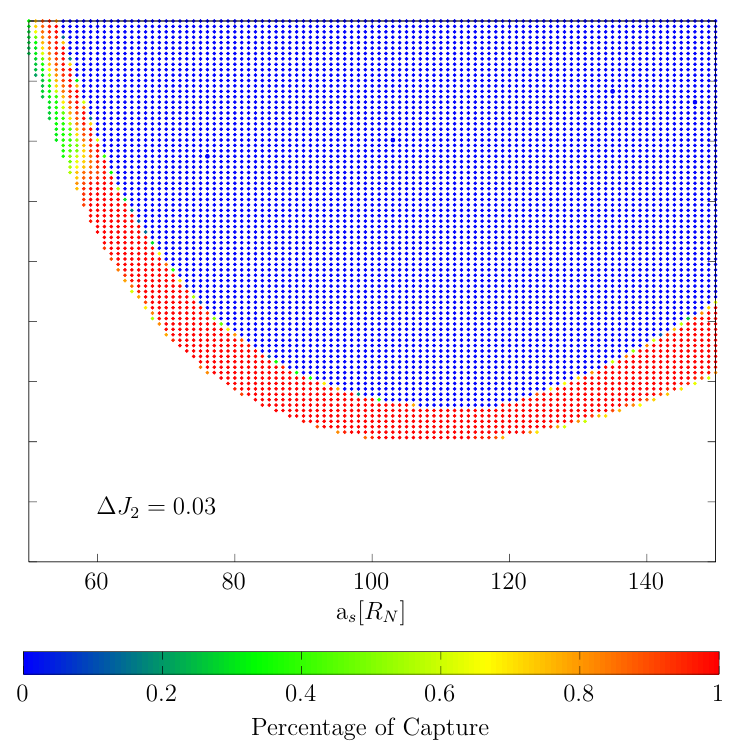}
    \includegraphics[width=0.31\textwidth]{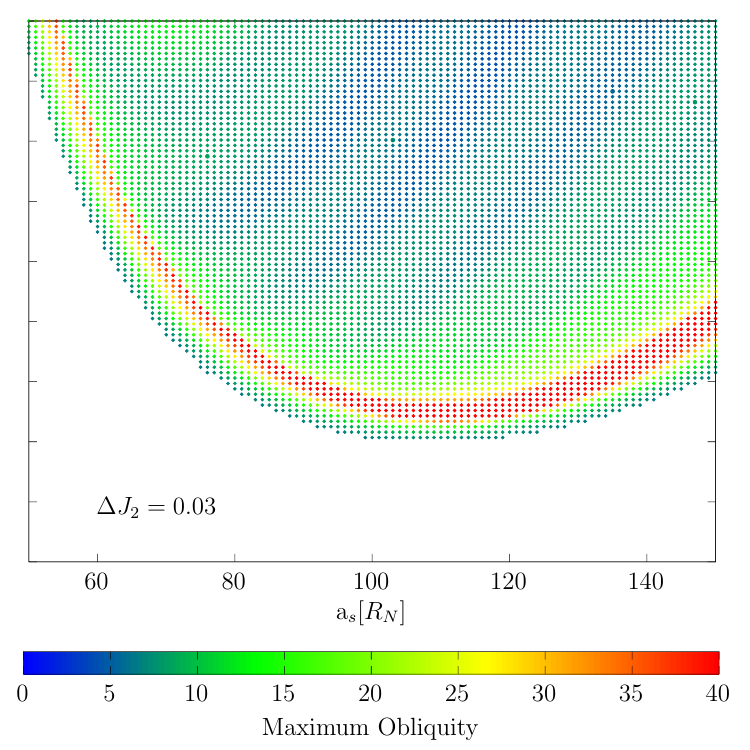}
    \caption{Distribution  of percentage of success (first column), capture into resonance (second column) and maximum obliquity reached (third column) in the parameter space $\mathcal{S}$. Each row corresponds to a fixed value of $\Delta J_2$, from top to bottom: $0.00$, $0.01$, $0.02$, $0.03$. The color bar indicates the value of the quantity represented.}
    \label{fig:fullpage_myplot}
\end{figure*}

\subsection{Conditions for successful tilting of Neptune}\label{ssec_discres}

Exploring the parameter space as described in Sect~\ref{ssec-stats}, we found that simulations that lead to a successful tilting of Neptune are clustered in a thin band, characterized by moderate values of the final precession constant. Their distribution in the parameter space $\mathcal{S}$ is shown in \figurename~\ref{fig:fullpage_myplot}. 

The relatively narrow range of parameters that produces successful runs is the result of the interplay between two main factors: the probability of capture into resonance during migration and the efficiency of obliquity excitation. The first one decreases for larger values of the final precession constant $\alpha$. This is due to the fast rate with which the resonance is crossed (see \figurename~\ref{fig:migration_case1}). Since capture into resonance requires a variation timescale of $\alpha / |s_8|$ lower or comparable to the libration period within the resonance, if $\alpha$ is too large, the obliquity experiences just a kick at the crossing of the resonance. Apart from rare cases positioned near the upper boundary of the success region, this kick is not sufficient to tilt Neptune up to $28.3^\circ$. The second factor, i.e. efficiency of obliquity excitation, is also influenced by $\alpha$. While captured into resonance, Neptune's obliquity increases up to a maximum value that depends on the final value of $\alpha / |s_8|$. If the ratio surpasses $1$ only of a small amount, the maximum obliquity does not manage to reach  $28.3^\circ$. The result is a thin region in the parameter space where both conditions are met leading to successful excitation. 

From \figurename~\ref{fig:fullpage_myplot}, we can appreciate how the results depend weakly on the initial conditions chosen for Neptune's spin-axis: success and capture rates different from $0$ or $1$ are only found next to the boundary of the region described above. This makes our results quite general, as they do not require specific initial conditions of Neptune's obliquity (taken from $1$ to $5$ degrees) and spin-axis orientation (taken from $0$ to $2\pi$ radians).

\subsection{Conditions for satellite survival}

\cite{Clement-etal_2026} showed that in case of violent instabilities during planets' evolution, one or multiple close encounters between the giant planets can occur, threatening the integrity of their satellite systems. In particular, they found that close encounters as deep as $0.03$ au ($\approx 176\,R_U$) between Uranus and another ice giant is sufficient to completely destabilize the Uranian satellite system. Similar results have been obtained in \cite{Deienno-etal_2014} for the Jovian system, where encounters closer than $0.02$ au ($\approx 42.8\,R_J$) can disrupt the orbits of the regular satellites of Jupiter.

In Sect.~\ref{sec_5b}, we obtained similar results for the primordial satellite system of Neptune, with upper bound on $r_\text{peri}$ depending on the particular primordial satellite system configuration. For example, taken an outer moon with $a_s = 100\, R_N$ and $m_s = 3.5\times 10^{-4}\, M_N$, a single close encounter with an ice giant at a distance of $0.046$ au ($\approx 270\,R_N$)  or less can cause large changes in the semi-major axis and eccentricity of the outer satellite, or even its ejection. Encounters at larger distances do not necessarily disrupt the system, but can still significantly alter the satellite's orbit. Such changes may lead to destabilization following multiple encounters or interfere with the subsequent tilting process. 

To quantify the stability of the satellite system following an encounter, we consider the system to remain stable provided that $\Delta e< 0.05$ and $\Delta \ln\alpha < 0.05$. These criteria ensure that the satellite retains a relatively low eccentricity and that the change in the precession constant $\alpha$ remains moderate, thereby preserving the conditions required for maintaining the spin-orbit resonance. The conditions obtained can be rewritten in terms of $\Delta e$ and $\Delta a_s$, with the letter depending on its initial value. In our three examples, the condition on $\Delta \ln\alpha $ rewrites as $\Delta a_s < 2\, R_N$ if $a_s = 55\, R_N$, and $\Delta a_s < 5\, R_N$ if $a_s = 100\, R_N$ or $a_s = 145\, R_N$. These conditions are generally satisfied for encounters between an ice giant and Neptune with a distance larger than $0.039$ au ($\approx 230\, R_N$) if $a_s = 55\, R_N$, larger than $0.076$ au ($\approx 450\, R_N$) if $a_s = 100\, R_N$ and larger than $0.102$ au ($\approx 600\, R_N$) if $a_s = 145\, R_N$.

Since at the time of planetary instability the tilting of Neptune was not started yet, our scenario favors a relatively mild giant-planet instability, in order to avoid extremely deep mutual encounters. While this requirement may exclude the most violent versions of the Nice model, it is compatible with other dynamical constraints, such as the preservation of the Galilean satellites' Laplace resonance, and satellite-assisted tilting scenarios for Uranus \citep{Boue-Laskar_2010,Deienno-etal_2014,Saillenfest-etal_2022}.

In the dynamical pathway that we proposed in this work, the tilting of Neptune is generally completed mid way through the planetesimal disk crossing phase. After Neptune was successfully tilted, the outer satellite can be destabilized and ejected by a single or multiple close encounters with sufficiently massive planetesimals, as discussed in Sect.~\ref{ssec_planetesimal}. This mechanism is not just possible, but also necessary to reconcile our tilting scenario with the present-day configuration of the Neptunian system, as we discuss in the next section.

\subsection{Exit from spin-orbit resonance}
The system produced at the end of our successful simulations still contains a massive and distant satellite, and remains trapped in the spin-orbit resonance with the nodal frequency $s_8$ (see Figure~\ref{fig:migration}). By contrast, the current Neptunian system is characterized by the absence of such a satellite and by a non-resonant spin-axis state (see Table~\ref{tab:obliquities}). Moreover, Neptune hosts Triton, a single massive moon on a retrograde and highly inclined orbit. 

Triton is thought to have been captured during Neptune's migration through the trans-Neptunian planetesimal disk, likely when the planet was located between $\sim 20$ and $30$ au \citep{Agnor-Hamilton_2006,Nesvorny-etal_2007}. Although the exact capture mechanism remains debated \citep{Agnor-Hamilton_2006,Nogueira-etal_2011,Rufu-Canup_2017}, the presence of a massive and distant regular satellite at the time of capture would probably strongly interfere, or even prevent, Triton's successful insertion and long-term survival. This suggests that Neptune’s primordial outer satellite must have been removed prior to Triton’s capture.

A natural mechanism for this removal is suggested in Section~\ref{ssec_planetesimal}, where we showed that the outer moon could be ejected during a sufficiently close encounter with a planetesimal. In the scenario considered here, the conditions required for such an event are rather stringent, particularly because the intruder must be relatively massive. A more detailed study about the effect of planetesimals on satellite systems of giant planets is still necessary (see also \citealp{Clement-etal_2026} in the case of Uranian moons).
Nevertheless, we provide here an order-of-magnitude estimate of the probability of ejection in our scenario, in the case where $a_s = 100\, R_N$. If the two orbits are coplanar, the outer moon and the intruding planetesimal can approach each other closely provided that their orbits intersect. This condition is satisfied if and only if $r_\text{peri}<a_s$. To treat the problem approximately as planar, we require the inclination of the encounter orbit to lie within a narrow range around that of the moon, namely $i_\text{enc}\in(i_s-\delta,i_s+\delta)$. Furthermore, the intruding planetesimal must be both relatively massive, with $m_{\text{enc}}>m_s$, and slow, with $v_\infty<3.5,\mathrm{km,s^{-1}}$.

Supposing that the three request are independent, we define the probability of such an encounter as
\begin{align}
    \mathcal{P}(\text{encounter}) &= \mathcal{P}(A)\times \mathcal{P}(B)\times \mathcal{P}(C)\times \mathcal{P}(D)
\end{align}
where $A = \{r_p\in(40,100)\, R_N\}$, $B = \{v_\infty<3.5\, \text{km s}^{-1}\}$,  $C = \{m_\text{enc}>m_s\}$, and $D = \{i_\text{enc}\in (i_s-\delta,i_s+\delta)\}$. Here $\mathcal{P}(A)$ and $\mathcal{P}(B)$ can be estimated from Fig.~2 of \cite{Nogueira-etal_2011}, while the probability associated with the mass range depends on the mass distribution of the planetesimal disk. Since the latter is poorly constrained, we denote it by $q_m$. Finally, the outer moon lay on its Laplace plane with $i_s \approx \epsilon/2$ since $a_s\approx r_M$. For $\mathcal{P}(D)$ we will use the current inclination distribution of KBO as an approximation \citep{Gulbis-etal_2010}.
However, satisfying these conditions does not necessarily guarantee the ejection of the outer moon. For ejection to occur, the planetesimal must also pass sufficiently close to the moon. Let us assume that ejection takes place whenever the mutual distance between the outer moon and the intruder is less or equal than $d_\text{max}$. Assuming that the orbital phase of the moon is uniformly distributed, the probability of finding the moon within this distance when the planetesimal reaches conjunction is
\begin{equation}
    \mathcal{P}(\text{ejection}|\text{encounter}) \approx 2\frac{2d_\text{max}}{2\pi a_1}= 2\frac{d_\text{max}}{\pi a_1} \,.
\end{equation}

The value of $d_{\max}$ depends on the properties of the encounter, namely $(r_{\text{peri}},v_\infty,m_{\text{enc}})$, and is expected to range between $0.2$ and $1\,R_N$. A proper determination of this quantity would require an extensive suite of numerical simulations, which is beyond the scope of the present work. As a first approximation, we therefore adopt a representative average value of $\overline{d_{\max}}=0.5\, R_N$. The probability of ejection during a single encounter can then be estimated as
\begin{align}
    \mathcal{P}(\text{ejection}) &= \mathcal{P}(\text{ejection}|\text{encounter})\times \mathcal{P}(\text{encounter}) \notag\\
     & \approx 2\frac{\overline{d_\text{max}}}{\pi a_1}\times  1.2\times10^{-3}\times 0.85\times q_m \times10^{-3} \ \notag\\
     &\approx 2\frac{\overline{d_\text{max}}}{\pi a_1}\times10^{-6}\times q_m \approx 3.1\times 10^{-9}\times q_m\,.
\end{align} 
For the computation of $\mathcal{P}(D)$ we chose $\delta = 0.3^\circ$ so that the distance between the outer satellite orbit and the encounter orbital plane is at most $\overline{d_\text{max}}$.

The probability $\mathcal{P}(\text{ejection})$ is extremely small for an individual encounter, mainly because of the stringent geometrical and mass requirements. Nevertheless, one should also account for the enormous number of encounters experienced by Neptune, estimated to be $n\approx 1.7\times10^6$ \citep{Nogueira-etal_2011}. Assuming statistically independent encounters, the probability of at least one successful ejection event is
\begin{align}
    \mathcal{P}(N_\text{ejection}\geq1) = 1-(1-\mathcal{P}(\text{ejection}))^n\,.
\end{align}

This probability depends on the poorly constrained mass distribution of the planetesimals. For instance, for $q_m=0.01$, we obtain $\mathcal{P}(\text{ejection})\approx 0.005\%$, whereas for $q_m=0.1$, we obtain $\mathcal{P}(\text{ejection})\approx 0.05\%$. We emphasize that this estimate is only approximate and that a more reliable assessment would require numerical simulations accounting for the detailed dynamics of the encounters and the mass distribution of the planetesimals. In any case, the resulting probability appears very low, making the ejection of the outer moon due to a single encounter with a planetesimal an exceptional (but not impossible) event.

\begin{figure}
    \centering
    \includegraphics[width=1.0\linewidth]{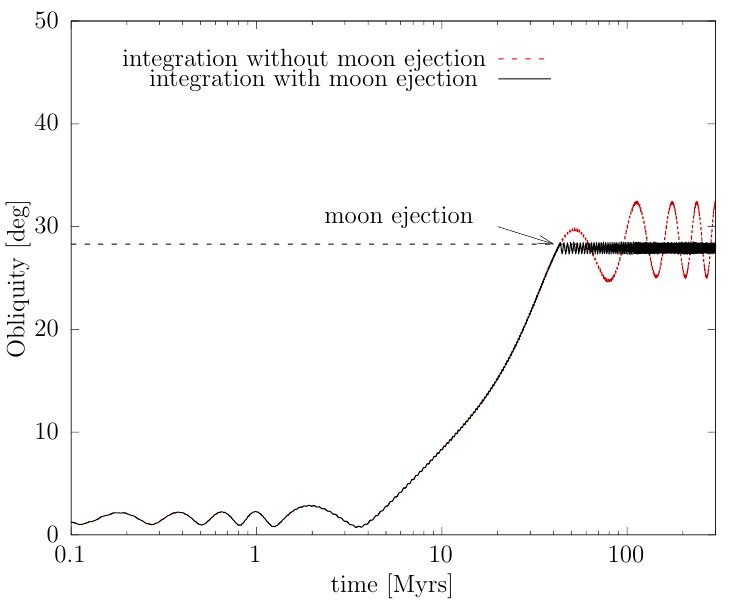}
    \caption{Evolution of Neptune's obliquity with (black line) and without (red dashed line) the ejection of the outer moon.}
    \label{fig:triton_capture}
\end{figure}

The loss of the outer satellite provides a natural mechanism for escaping the spin-orbit resonance. To investigate the consequences of such an event, we considered successful tilting simulations and artificially removed the outer satellite once Neptune's obliquity reached $28.3^\circ$, mimicking its loss through a close encounters. The resulting evolution of Neptune's spin axis is shown in Fig.~\ref{fig:triton_capture}. Prior to the removal, Neptune is trapped in resonance and its spin axis librates around the Cassini state C2 . The sudden loss of the outer satellite reduces the satellite-induced contribution to the precession constant $\alpha$, causing the ratio $\alpha/|s_8|$ to drop below unity. As a consequence, the resonant equilibrium disappears and the spin axis passes from libration to circulation. Although the obliquity undergoes oscillations during this transition, it remains close to the value reached at resonance breaking, thereby preserving the large tilt.

\begin{figure}[h!]
    \centering
    \includegraphics[width=1.0\linewidth]{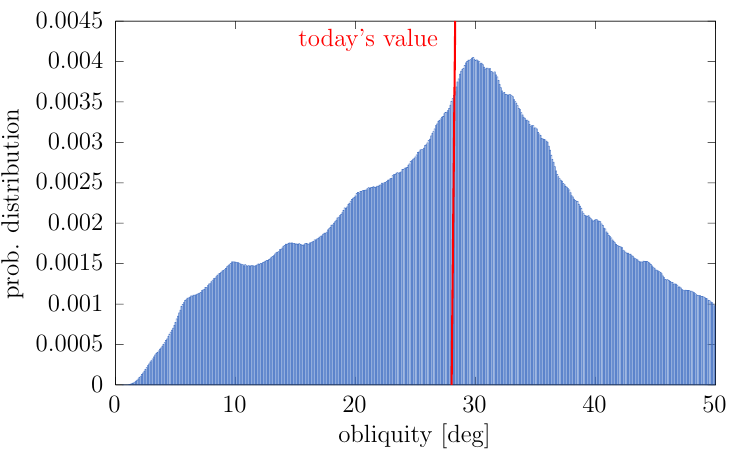}
    \caption{Distribution of Neptune's obliquity at the time of planetesimal encounter and consequent exit from the resonance.}
    \label{fig:freq_ob}
\end{figure}

The exact value of $28.3^\circ$ is therefore not deterministic but rather stochastic, arising from the wide libration of the spin axis around the equilibrium point C2 and the precise epoch of the outer moon ejection. Nevertheless, taking for example all successful simulations with $\Delta J_2=0.03$ (see \figurename.~\ref{fig:fullpage_myplot}), we computed the distribution of Neptune's final obliquity considering that the exit from resonance could have occurred at any time during the crossing of the planetesimal disk. Such a distribution results to be peaked around $30^\circ$ (see \figurename~\ref{fig:freq_ob}), making $28.3^\circ$ a plausible value for Neptune's obliquity for the scenario we presented.

\subsection{Implications on Triton and Nereid}

After the ejection of the outer satellite, the remaining system would have consisted of one (or more) inner moon with a total mass of order $5\times10^{-5}\, M_N$ located at $\sim 35\, R_N$. This configuration is perfectly consistent with the conditions explored by \citet{Rufu-Canup_2017}, who showed that the presence of a low-mass inner satellite system increases the likelihood of Triton's successful capture and subsequent fast evolution toward its present orbit. As, in this scenario, the timescale of circularization of the orbit of Triton was only $10^5$ years \citep{CUK-GLADMAN_2005,Rufu-Canup_2017}, even though Neptune crossed again the spin-orbit resonance with $s_8$ \citep{Gomes_2026}, capture into resonance would have been not possible. Therefore, we expect that the following orbital evolution of Triton had only a minor impact on the obliquity of the planet. In the end, this sequence — tilting through spin-orbit resonance, destabilization and loss of the outer satellite, and subsequent capture of Triton — provides a coherent pathway linking the primordial Neptunian system to its present-day architecture.

Another important aspect to consider when investigating the evolutionary history of the Neptunian satellite system is the presence of the outer moon Nereid on its peculiar, highly eccentric orbit ($a=224\, R_N$, $e=0.75$). Nereid origin is still debated, being either a captured irregular satellite or a primordial moon of Neptune. Previous works showed that the capture and subsequent evolution of Triton driven by tides would have destabilized Nereid, while a fast circularization of Triton through collisions with other moons could have allowed to preserve it.

In our scenario, Nereid could be either captured after the outer moon ejection or be a primordial satellite of Neptune. However, in the second case, its outer orbit could have been extremely vulnerable to encounters with planetesimals, so that it is difficult to account for both the ejection of the outer massive moon and the preservation of Nereid on its current orbit.

In this context, \cite{Belyakov-etal_2026} recently proposed a new possible explanation for Nereid's orbit. They found that, based on its surface composition, Nereid was likely a regular satellite of Neptune. They then proposed that a massive planetesimal may have disrupted Neptune's primordial satellite system, leaving Nereid on its present-day eccentric orbit. After this event, Triton was captured and evolved rapidly to its current orbit as described in \cite{CUK-GLADMAN_2005}. The two-step evolution depicted by \cite{Belyakov-etal_2026} is consistent with our scenario, even though they did not need to include a large outer moon in their model. However, we showed that such a moon could be the missing ingredient to explain the current tilt of Neptune without invoking giant impacts.

%% file: sections/07_Conclusions.tex
\section{Conclusion}\label{sec_conclusions}

Spin-orbit resonances have been shown to be a viable candidate for tilting the giant planets of the solar system, like Jupiter \citep{Ward-Canup_2006,Saillenfest-eatal_2020,Dbouk-Wisodm_2023}, Saturn \citep{Ward-Hamilton_2004,Saillenfest-etal_2021a,Wisdom-etal_2022}, and Uranus \citep{Rogoszinski-Hamilton_2020,Saillenfest-etal_2022}.

Applying this tilting mechanism to Neptune is challenging due to its current spin-axis precession frequency, which is much smaller than any orbital frequency of the solar system. In fact, the presence of Triton on its current orbit, the only large satellite of Neptune, does not boost enough the precession rate of the planet. The recent work of \citet{Gomes_2026} showed that a slow circularization of an highly eccentric orbit of Triton could have made Neptune trigger a spin-orbit resonance with $s_8$, successfully tilting the planet.

This mechanism fails in the case of a very fast circularization and orbital evolution as predicted by \citet{CUK-GLADMAN_2005}. However, if in the past Neptune had hosted a primordial satellite system, it is possible that Neptune would have triggered the spin-orbit resonance with $s_8$ during the planetary migration.

In this work, we determined the properties of a primordial satellite system that would have enabled Neptune to encounter the spin-orbit resonance associated with $s_8$. We considered a two-satellite configuration consisting of an inner moon that increases Neptune's effective quadrupole moment by an amount $\Delta J_2$ and an outer moon of mass $m_s$ orbiting at a semi-major axis $a_s$. The inner satellite reduces the mass required for the outer satellite to drive Neptune across the resonance. This two-satellite representation is not restrictive: a system containing several suitably distributed moons could produce the same effective $\Delta J_2$ and the same total increase in the spin precession constant $\alpha$.

We then defined a region of the parameter space of the primordial satellite system for which Neptune could have crossed the spin-orbit resonance with $s_8$ during the planetary migration. We sampled this region and we ran thousands of numerical simulations of the evolution of the spin-axis orientation of Neptune, starting from low initial obliquities ($\epsilon \lesssim 5^\circ$). The outcomes of these simulations highlighted the presence of an open set in the parameter space for which Neptune's obliquity would have been excited up to its current value. In particular, the total mass of the primordial satellite system is required to be $\geq 4\times 10^{-4}$. As an example, for $\Delta J_2=0.03$ (realized for example by an inner moon of mass $5\times10^{-5}\, M_N$ and semi-major axis $35\,R_N$) and an outer satellite with mass $m_s\approx 3.5\times 10^{-4}\, M_N$, and semi-major axis between $100$ and $120\ R_N$, we showed that Neptune would have been captured into resonance and its obliquity would have increased up to $28^\circ$ (see \figurename~\ref{fig:fullpage_myplot}). A variety of configurations with larger masses of the moons are possible, but are less favored by formation models.

Subsequently, we investigated the effects of gravitational encounters between Neptune and Uranus during the planetary migration on the primordial satellite system. We found that the latter could have survived if the encounters were not too close, with a minimum distance larger than $0.076$ au. This condition is compatible with Nice model simulations that are not too chaotic and avoid violent orbital crossings between the giant planets \citep{Gomes-etal_2005,Tsiganis-etal_2005,Morbidelli-etal_2007}.

Once Neptune had reached its present obliquity, the long-term stability of the primordial satellite system was no longer required. One or more subsequent close encounters with massive planetesimals from the planetesimal disk could have destabilized the primordial satellite system, leading to the ejection of the outer moon. The dynamical conditions for the removal of the outer moon resulted to be quite stringent, so that the overall feasibility of this process needs to be assessed in the future through a dedicated investigation, possibly involving also the process that led to the capture of Triton. Nevertheless, this event could have acted both as a mechanism to escape from the spin-orbit resonance (see also \citealp{Wisdom-etal_2022} for a similar event in the Saturn system), freezing the obliquity of Neptune to its current value, and as a preparatory step for the capture of Triton. 

After the ejection of the outer moon, the primordial satellite system architecture matched the dynamical conditions required for eventually reaching the current configuration of the moon system of Neptune. More precisely, \citet{Rufu-Canup_2017} showed that, if Neptune had hosted a primordial satellite system with a mass between about $0.3$ and $1\times 10^{-4}\, M_N$, Triton would have destabilized the primordial moons and would have rapidly evolved toward its current orbital state, preserving also the orbit of Nereid.

In conclusion, we showed that a primordial satellite system with a total mass of around $4\times 10^{-4}\,M_N$ is sufficient to tilt Neptune during planetary migration. Even though the current satellite system of Neptune is less massive, this value is about $2$ or $4$ times larger than what observed in Jupiter and Saturn's system, and what predicted by models of satellite formation around giant planets \citep{Mosqueira-Estrada_2003,Canup-Ward_2006}. Moreover, it is comparable to the mass of the moon required to tilt Uranus without giant impacts \citep{Saillenfest-etal_2022}.